\documentclass[onecolumn,aps,pra,preprintnumbers,amsmath,amssymb]{revtex4}

\usepackage{natbib}
\usepackage{graphicx}
\usepackage{bm}
\usepackage{color}

\usepackage{float}
\usepackage{amsmath}
\usepackage{amsfonts}
\usepackage{amssymb}
\usepackage{makeidx}
\usepackage{graphicx}
\usepackage{placeins} 
\usepackage{color}
\usepackage{physics}

\begin{document}
	
	\title{ Non-Markovianity in a dressed qubit with local dephasing}
	\author{Saima Bashir$^1$}
	
	\author{Muzaffar Qadir Lone$^2$ \footnote{corresponding author: lone.muzaffar@uok.edu.in}}
	\author{Prince A Ganai$^3$}
	\affiliation{ $^1$ Department of Physics, National Institute of Technology, Srinagar-190006 India \\
		$^2$ Quantum Dynamics Lab, Department of Physics, University of Kashmir, Srinagar-190006 India}
	
\begin{abstract}
	We study the dynamics of a dressed qubit implemented by a spinless fermion hopping between  two lattice sites with each site strongly coupled to a bath of phonons. We employ Lang-Firsov  transformation to make the problem tractable perturbatively. Applying time-convolutionless master equation within the polaron frame, we investigate decoherence dynamics of the dressed qubit within the singlet- triplet basis of the system for a wide range of bath spectral densities. It is shown that the coherence persists for longer time scales for large coupling values and shows non-monotonic behaviour reflecting the presence of non-Markovianity in the dynamics. Non-Markovianity, characterized by coherence revivals and non-monotonic decay patterns, emerges distinctly depending on the bath spectrum and coupling strengths. Systems coupled to sub-Ohmic baths, whether both or in combination with another type, display pronounced memory effects at relatively small values of couplings. In contrast, combinations involving Ohmic and super-Ohmic baths exhibit noticeable non-Markovianity only at higher couplings.
\end{abstract}
	\maketitle
	\section{Introduction}
A realistic quantum system is inherently open, as it unavoidably interacts with a  bath  that
leads to the loss of quantum coherence through a process known as decoherence \cite{breuer2002theory,Weiss2012,leggett1987dynamics}. This phenomenon not only degrades the coherence of quantum states but is also often accompanied by energy dissipation and information loss, posing significant challenges to the practical realization of quantum technologies. As such, understanding, mitigating, and controlling decoherence is crucial for the advancement of quantum information processing, communication, and storage devices~\cite{horodecki2009quantum,nielsen2010quantum,pirandola2015advances,cozzolino2019high,guo2019advances}. Beyond its technological implications, decoherence also holds fundamental importance
in the study of quantum mechanics itself, as it offers insights into the transition from quantum to classical behavior and the emergence of classicality in macroscopic systems \cite{pirandola2015advances,leggett1987dynamics,yu2003}.

The concept of dressed qubits has emerged as a valuable framework for examining system-bath interactions at the microscopic level \cite{Lidar, Preskill,Ts}. Dressed qubits are quantum bits whose properties are renormalized through their coupling with environmental degrees of freedom. For example, polaron$-$where an electron is coupled to a lattice vibration; polaritons where matter excitations strongly interact with photons, etc. These coupled systems exhibit a rich interplay of coherent dynamics, dissipation, and decoherence offering a platform to probe fundamental aspects of open quantum systems \cite{Nazir,Lasek,Pichler,yu2003}. There are several models that provide a description of these dressed states. The Holstein model \cite{Macridin2004, DattaYarlagadda2007,RejaYarlagadda2012}, in particular, describes a tight-binding electron interacting locally with dispersionless optical phonons, making it a useful framework for studying local lattice distortions and their impact on electron dynamics \cite{Zheng,Klein2023,Chakraborty2023}.
	
Open quantum systems have advanced in recent years in context of quantum non-Markovianity, an effect that
reflects the back-action of the environment on a system’s dynamics \cite{Haikka,Li2018}. Unlike Markovian environments, where information irreversibly flows from the system to bath, non-Markovian environments \cite{Rivas2010,Breuer2016} can temporarily store and return information to the system. This backflow of information can partially restore lost coherence or correlations and is typically manifested through a departure from monotonic decay in certain dynamical quantities \cite{Liu2011,Titas}. Several theoretical and experimental studies have examined the impact of non-Markovian, structured environments on quantum coherence, correlations, and entanglement\cite{Gan, Chaing}. Various quantitative measures of non-Markovianity like CP-divisibility \cite{Rivas2010}, distance-based measures \cite{BLP}, coherence-based indicators using the $l_1$ norm of coherence \cite{Titas} etc, are based on information backflow, that is any deviation from strictly monotonic behaviour signals non-Markovian effects in the dynamics. 
	
In this work, we analyze  dynamics of a qubit modeled by a spinless fermion hopping between two lattice sites, wherein each site is strongly coupled  to a bath of phonons \cite{Dutta,Datta,Lone}. 
The bath at each site is characterized by spectral density of the form $ \omega^s e^{-\omega^2 / \Omega^2}$ with a cutoff $\Omega $ and $s\ge 0$ allowing to choose different scenarios: $s=0.5 ({\rm sub-Ohmic}), s=1 ({\rm Ohmic}),s=2({\rm super-Ohmic})$. { There are several works that deals with dephasing in such dressed qubits \cite{Lidar,Preskill,Ts,mac,saima,Lone}, however either the local dephasing effects are not taken into account or the system-bath interactions are assumed to be weak or the bath at each site is same. For example, Lidar et. al \cite{Lidar} and Machinkowski et. al \cite{mac} introduced the concept of dressed qubits in solid state qubit architectures, and demonstrated their robustness against noise. However, their work is limited and does not incorporate explicitly non-Markovinity induced by baths. Furthermore, Konstantinos et. al. \cite{Ts} experimentally implemented singlet-triplet dressed qubits in semiconductor germanium. Their results indicate the extending coherence times by external control mechanism. However, the memory effects that could potentially arise are not taken into account 
as no structured bath is assumed. In contrast, our work systematically investigates the dynamics of a dressed qubit realized as a spinless fermion hopping between two lattice sites, each coupled to independent phonon baths. By incorporating strong system-bath couplings and a wide range of bath spectral densities, we demonstrate the emergence of long-lived coherence and non-Markovian dynamics, highlighting the interplay between bath exponents and site-specific couplings.
}

 To simplify the analysis and capture the bath-induced renormalization effects, we employ a unitary transformation known as Lang-Firsov transformation, \cite{Lang, Tara, Mahan,Cui} that eliminates the explicit system-bath interaction by embedding the bath effects into renormalized hopping terms and site energies. Applying quantum master equation, we solve for the coherence and population dynamics of the dressed qubit in the singlet-triplet basis. The non-monotonic behavior of coherence under temporal evolution signals non-Markovian effects. These effects are quantified through a non-Markovinity measure based on the coherence\cite{Titas,Cramer}. It is shown that different bath combinations at two sites show distinct non-Markovian behaviour.

This paper is structured in the following way. Section II describes model calculations. Here we introduce our
model and find effective system-bath interaction Hamiltonian. Section III deals with dynamical analysis of coherence in singlet-triplet basis with different baths at two sites. The non-Markovinity of the dynamics is analyzed through coherence measure in section IV. Finally, we conclude in section V.
	
	\section{MODEL FOR LOCAL PHONON COUPLING}
\begin{figure}[t]
		\includegraphics[width=4.5cm,height=4.25cm]{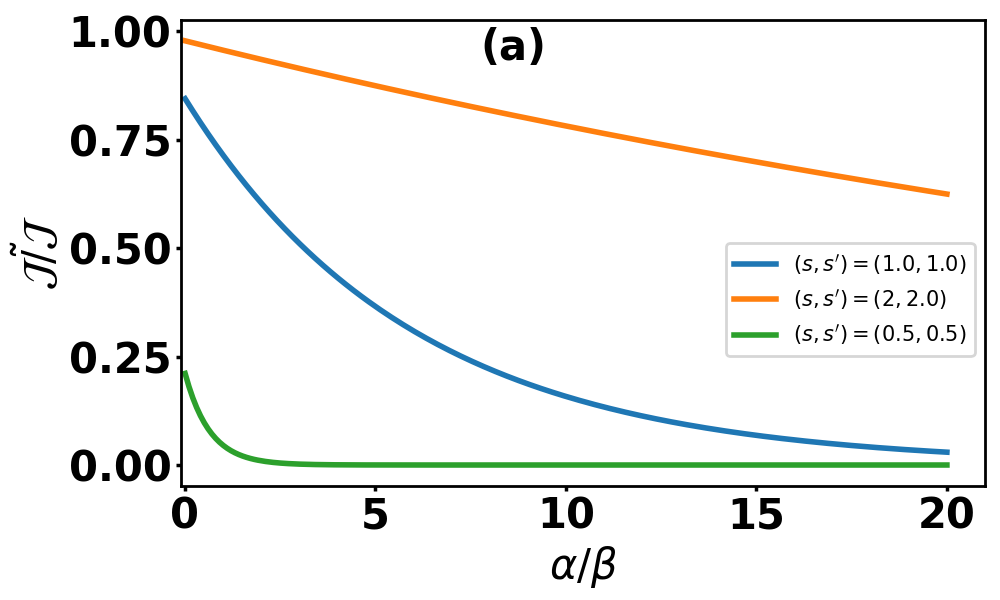} \hspace{0.2cm}
		\includegraphics[width=4.5cm,height=4.25cm]{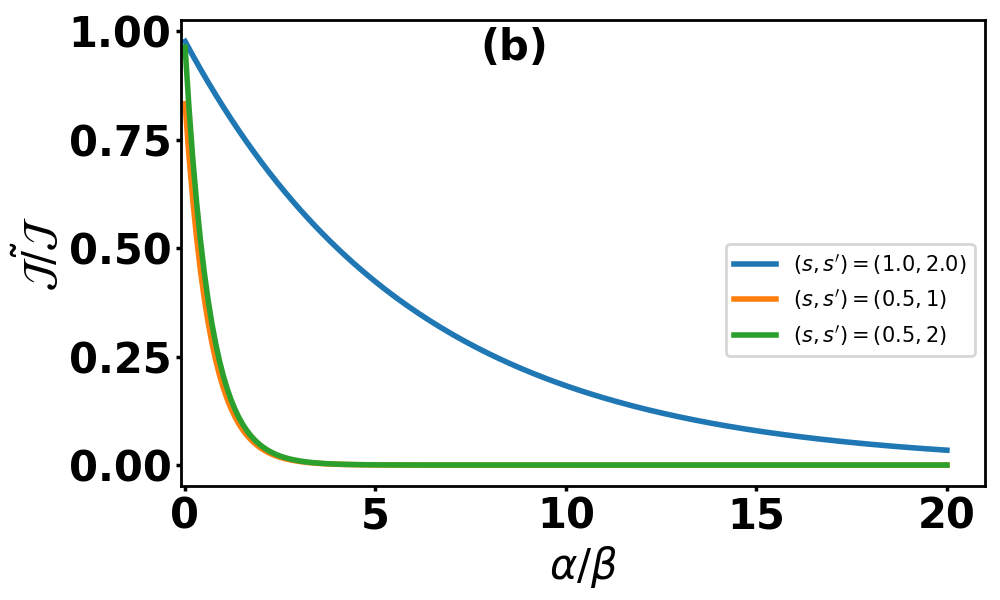}\hspace{0.2cm}
		\caption{{Variation of the renormalized hopping amplitude, $\frac{\tilde{\mathcal{J}}}{\mathcal{J}}$, as a  function of relative coupling strengths $\alpha$ and $\beta$ for the two sites, each coupled to independent phonon baths with spectral densities $J_1(\omega) = \alpha \, \omega^s e^{-\omega^2/\Omega^2}$ and $J_2(\omega) = \beta \, \omega^{s'} e^{-\omega^2/\Omega^2}$, (see section \ref{PP} for details). The bath exponents are chosen to represent different regimes: sub-Ohmic ($0<s,s'<1$), Ohmic ($s=s'=1$), and super-Ohmic ($s,s'>1$). We see that in all possible combinations, the effective coupling $\tilde{\mathcal{J}}$ is reduced in polaron frame.}}
		\label{SS}
	\end{figure}

    Here in this section, we examine a model in which a dressed qubit is realized as a spinless fermion hopping between two lattice sites, with each site strongly coupled to local phonon modes \cite{dey,Klein2023,Cao,saima}. The total Hamiltonian of the system is given by
	\begin{eqnarray*}
		H = H_{\text{fermion}} + H_{\text{bath}} + H_{\text{coupling}}.
	\end{eqnarray*}
    The behavior of a spinless fermion is governed by the tight-binding Hamiltonian
	\begin{eqnarray*}
		H_{\text{fermion}} = -\mathcal{J} (c_1^\dagger c_2 + c_2^\dagger c_1) + \sum_{i=1,2} \mu c_i^\dagger c_i,
	\end{eqnarray*}
Here, $c_i^\dagger$ and $c_i$ represent the creation and annihilation operators for a fermion at site $i$, respectively. The parameter $\mathcal{J}$ denotes the hopping amplitude between the two sites, while $\mu$ is the on-site energy. Each site is coupled to local phonons, decribed by the  Hamiltonian:

	\begin{eqnarray*}
		H_{\text{bath}} = \sum_{i=1,2} \sum_k \omega_{ik} a_{ik} ^\dagger a_{ik} ,
	\end{eqnarray*}
	where $a_{i,k}^\dagger$ and $a_{i,k}$ denote the creation and annihilation operators for the $k$-th bosonic mode at site $i$, and $\omega_{ik}$ represents the frequency of that mode. The interaction between the fermion and the phonons is described by:
	\begin{eqnarray*}
		H_{\text{coupling}} = \sum_{i=1,2} \sum_k  n_i (g_{ik} a_{ik} ^\dagger +g^*_{ik} a_{ik} ),
	\end{eqnarray*}
   where $g_{ik}$ characterizes the coupling strength between the $i$th site with $k$-th bosonic mode at that site. The operator $n_i = c_i^\dagger c_i$ denotes the number operator for fermions at site $i$.	
    This  type of model can be physically realized in various settings, including semiconductor quantum dots with electron-phonon interactions \cite{Brandes2005,Fujisawa1998}, superconducting qubits coupled to surface acoustic wave devices \cite{Gustafsson2014,Manenti2017}, trapped ions interacting with vibrational modes \cite{Blatt2012}, molecular junctions featuring electron-phonon transport \cite{Galperin2007}, and hybrid systems such as nitrogen-vacancy centers in diamond \cite{li}. These implementations make it a valuable framework for exploring quantum transport, coherence, and dissipation \cite{Damanet2019,Harrington2022}.
	Since the system-bath coupling is strong, it is useful to apply the unitary transformation that decouples system-bath interaction and such unitary transformation in this context is Lang-Firsov transformation \cite{Lang}. The Lang-Firsov transformation introduces a unitary transformation $ U = \exp\left(-\sum_{i,k} n_i (\frac{g_{ik}}{\omega_k} a_{ik} ^\dagger - \frac{g^*_{ik}}{\omega_k}  a_{ik} )\right) $, which shifts the bosonic operators and effectively diagonalizes the fermion-boson coupling term. Under this transformation, the Hamiltonian is rewritten as:
   \begin{eqnarray*}
	\tilde{H} &=& e^U	H e^{-U}= 	\tilde{H}_{\text{F}}+ 	\tilde{H}_{\text{B}}+ 	\tilde{H}_{\text{C}},
   \end{eqnarray*}
	where the transformed fermion Hamiltonian becomes
	\begin{eqnarray*}
		\tilde{H}_{\text{F}} = \tilde{\mathcal{J} }[c_1^\dagger c_2+c_2^\dagger c_1]+  \sum_{i=1,2} \tilde{\mu} c_i^\dagger c_i,
	\end{eqnarray*}
    where $\tilde{\mu}=\mu +\sum_{{i=1,2}}\sum_k\frac{|g_{ik}|^2}{\omega_{ik}}$ and  $\tilde{\mathcal{J} }=\mathcal{J}  e^{-\frac{1}{2}\sum_k(\frac{|g_{1k}|^2}{\omega^2_{1k}}+\frac{|g_{2k}|^2}{\omega^2_{2k}})}$ is the effective hopping and the transformed bosonic bath Hamiltonian becomes
	\begin{eqnarray*}
		\tilde{H}_{\text{B}} =  \sum_{i=1,2} \sum_k \omega_{ik} a_{ik} ^\dagger a_{ik}.
	\end{eqnarray*}
	The interaction Hamiltonian in the polaron frame becomes 
	\begin{eqnarray}
		\tilde{H}_{\text{C}}=	\tilde{\mathcal{J} }[\mathbb{B}c_1^\dagger c_2+\mathbb{B}^\dagger c_2^\dagger c_1],
	\end{eqnarray}
	where
\begin{eqnarray*}
	\mathbb{B}=e^{\sum_k\big(\frac{g^*_{1k}}{\omega_{1k}}a^ {\dagger}_{1k}-\frac{g^*_{2k}}{\omega_{2k}}a^ {\dagger}_{2k}\big)} e^{-\sum_k\big(\frac{g_{1k}}{\omega_{1k}}a_{1k} -\frac{g_{2k}}{\omega_{2k}}a_{2k} \big)}-1.
\end{eqnarray*}
	This transformation removes the direct qubit-bath coupling and instead encodes environmental effects through renormalized parameters $\tilde{\mathcal{J}}$ and bath operators $\mathbb{B}$.
{	In order to look at the behaviour of renormalized $\tilde{\mathcal{J}}$ with respect to the site-bath couplings, we introduce spectral density of each bath. 	 We define the coupling parameters $\alpha$ and $\beta$ as the strengths of the system-bath interaction for site 1 and site 2, respectively (see Section 3). Each site is coupled to an independent phonon bath, with spectral densities modeled as
		\begin{eqnarray*}
			J_1(\omega) = \alpha \, \omega^s \, e^{-\omega^2 / \Omega^2}, \quad
			J_2(\omega) = \beta \, \omega^{s'} \, e^{-\omega^2 / \Omega^2},
		\end{eqnarray*}
		where $\Omega$ is a high-frequency cutoff, and $s$ and $s'$ determine the type of bath: sub-Ohmic ($0<s,s'<1$), Ohmic ($s,s'=1$), or super-Ohmic ($s,s'>1$)}. In figure \ref{SS}, we plot renormalized coupling $\tilde{\mathcal{J}}$ with respect to relative coupling $\frac{\alpha}{\beta}$ for two sites. In all the different scenarios of bath combinations, we observe that the coupling ratio  $\frac{\tilde{\mathcal{J}}}{\mathcal{J}}$ is reduced in polaron or dressed basis, and {no divergences occur};  hence perturbation theory can be applied using $\tilde{H}_{\text{C}}$ as perturbation.  
	\section{Dynamics in polaron frame}\label{PP}
	We now analyze the dynamics of the qubit in dressed basis using  time-convolutionless (TCL) master equation approach. The TCL master equation is a powerful tool for studying open quantum systems, as it provides a systematic way to derive a time-local equation for the reduced density matrix of the system by effectively tracing out the bath degrees of freedom\cite{breuer2002theory,Clos2012, Shibata1977}. {In this work, we truncate the TCL expansion at second order in the system-bath coupling, which is sufficient to capture the essential decoherence and relaxation dynamics of the dressed qubit within the polaron frame.} Let $\rho_{\text{tot}}(t)$ represent the total density matrix for system plus bath. The reduced density matrix for the system is $\rho_S(t)=\mathrm{Tr}_B \rho_{\text{tot}}(t)$.
	The TCL formalism for reduced density matrix  $\rho_S(t)$ in the interaction picture, assuming bath density matrix to be $\rho_B$, is given by:
	\begin{eqnarray}
		\label{LE}
		\frac{d}{dt} \rho_S(t) = \int_0^t d\tau \, \mathrm{Tr}_B \Big( \big[ \tilde{H}_C(t), \big[ \tilde{H}_C(\tau), \rho_S(t)\otimes \rho_B \big] \big] \Big).
	\end{eqnarray}
	The operator $\hat{X}$ in the interaction picture is given by $\tilde{X}(t)=e^{iH_0t}\hat{X}e^{-iH_0t}$ so that $\tilde{H_C}$  can be written as $\tilde{H}_C(t) = e^{iH_0t}\tilde{H}_C e^{-iH_0t} $  with $H_0=\tilde{H}_{\text{F}}+ \tilde{H}_{\text{B}}$. After evaluating the double commutator in Eq.~\ref{LE}, we obtain the corresponding master equation (See appendix A for details)
	\begin{eqnarray}
		\label{KE}
		\frac{d{\rho_S}(t)}{dt} &=&{i \zeta(t)\bigg[ [(n_1-n_2)^2, \rho_S]\bigg] } + \Gamma_+(t) \bigg[\big\{c_2^{\dagger}c_1\rho_S c_1^{\dagger}c_2-\frac{1 }{2}\rho_S c_1^{\dagger}c_2c_2^{\dagger}c_1-\frac{1 }{2}c_1^{\dagger}c_2c_2^{\dagger}c_1\rho_S \big\}\bigg] \nonumber\\&&+\Gamma_+(t) \bigg[\big\{c_1^{\dagger}c_2\rho_S c_2^{\dagger}c_1- \frac{1 }{2}\rho_S c_2^{\dagger}c_1c_1^{\dagger}c_2-\frac{1 }{2}c_2^{\dagger}c_1c_1^{\dagger}c_2\rho_S \big\}\bigg] +\Gamma_-(t)\bigg[c_1^{\dagger}c_2\rho_S c_1^{\dagger}c_2+c_2^{\dagger}c_1\rho_S c_2^{\dagger}c_1\bigg]  
	\end{eqnarray}
	Where the functions representing different decay channels are
	\begin{align}
\Gamma_{\pm}(t)
&= 2\tilde{\mathcal{J}}^2 \int_0^t d\tau \Bigg[
e^{\pm \sum_k \Big(
\frac{|g_{1k}|^2}{\omega_{1k}^2} \cos [\omega_{1k} (t-\tau)]
+ \frac{|g_{2k}|^2}{\omega_{2k}^2} \cos [\omega_{2k} (t-\tau)]
\Big)} \nonumber \\
&\quad \times
\cos \Bigg(
\sum_k \frac{|g_{1k}|^2}{\omega_{1k}^2}
\sin [\omega_{1k} (t-\tau)]
+ \frac{|g_{2k}|^2}{\omega_{2k}^2}
\sin [\omega_{2k} (t-\tau)]
\Bigg)
-1
\Bigg],
\\[1ex]
\zeta(t)
&= \tilde{\mathcal{J}}^2 \int_0^t d\tau \Bigg[
e^{\sum_k \Big(
\frac{|g_{1k}|^2}{\omega_{1k}^2} \cos [\omega_{1k} (t-\tau)]
+ \frac{|g_{2k}|^2}{\omega_{2k}^2} \cos [\omega_{2k} (t-\tau)]
\Big)} \nonumber \\
&\quad \times
\sin \Bigg(
\sum_k \frac{|g_{1k}|^2}{\omega_{1k}^2}
\sin [\omega_{1k} (t-\tau)]
+ \frac{|g_{2k}|^2}{\omega_{2k}^2}
\sin [\omega_{2k} (t-\tau)]
\Bigg)
\Bigg].
\end{align}

The standard basis for the underlying qubit is $\{ c_1^\dagger|0\rangle, c_2^\dagger|0\rangle \}$ with $ \ket{0}$ representing an empty site and $\ket{1}=c^\dagger\ket{0}$ as the filled site. In this basis, we define two states of relevance called as singlet $\mathcal{S}$ and triplet $\mathcal{T}$ states:
    \begin{eqnarray*}
	|\mathcal{S}/\mathcal{T}\rangle = \frac{1}{\sqrt{2}} \big[ |10\rangle \pm |01\rangle \big].
   \end{eqnarray*}
These states evolve according to the master equation \ref{KE} as	
\begin{eqnarray}
	\label{DFE}
	\frac{d}{dt}\langle \mathcal{T}|{\rho_S(t)}|\mathcal{T}\rangle &= & -(\frac{2\Gamma_+(t) -\Gamma_-(t)}{2}) [\langle \mathcal{T}|{\rho_S}(t)|\mathcal{T} \rangle-\langle \mathcal{S}|{\rho_S}(t)| \mathcal{S} \rangle]  \\  	
	\label{DF}
	\frac{d}{dt}\langle \mathcal{S}|{\rho_S(t)}|\mathcal{S} \rangle &= &-(\frac{2\Gamma_+(t) -\Gamma_-(t)}{2})[\langle \mathcal{S}|{\rho_S}(t)|\mathcal{S} \rangle-\langle \mathcal{T}|{\rho_S}(t)|\mathcal{T}\rangle]
	\\
	\label{D}
	\frac{d}{dt}\langle \mathcal{T}|{\rho_S(t)}|\mathcal{S}\rangle &= &-(\frac{\Gamma_-(t)+6\Gamma_+(t)}{2})\langle \mathcal{T}|{\rho_S}(t)|\mathcal{S}\rangle - (\frac{\Gamma_-(t)-2\Gamma_+(t)}{2})\langle \mathcal{S}|{\rho_S}(t)|\mathcal{T}\rangle.
\end{eqnarray}
    By solving equations~\ref{DFE}, \ref{DF}, and \ref{D}, we obtain the time evolution of both the diagonal and off-diagonal elements as follows:
	\begin{eqnarray*}
		\langle \mathcal{S}|\rho_S(t)|\mathcal{S}\rangle &=& \frac{1}{2}\langle \mathcal{S}|\rho_S(0)|\mathcal{S}\rangle[1+\exp[-\int_0^t ds \Gamma_0(s)]] +  \frac{1}{2}\langle \mathcal{T}|\rho_S(0)|\mathcal{T}\rangle[1-\exp[-\int_0^t ds \Gamma_0(s)]] \\
		\langle \mathcal{S}|\rho_S(t)|\mathcal{T}\rangle &=& \frac{1}{2}\langle \mathcal{S}|\rho_S(0)|\mathcal{T}\rangle [e^{- \int_0^t ds \Gamma_1(s)}+ e^{-\int_0^t ds \Gamma_2(s)} ] +  \frac{1}{2}\langle \mathcal{T}|\rho_S(0)|\mathcal{S}\rangle [e^{- \int_0^t ds \Gamma_1(s)}-e^{-\int_0^t ds \Gamma_2(s)} ],
	\end{eqnarray*}
	where $ \Gamma_0(t)=\frac{2\Gamma_+(t) -\Gamma_-(t)}{2} $, $\Gamma_1(t)= 2\Gamma_+(t)+ \Gamma_-(t)$ and  $ \Gamma_2(t)= 4\Gamma_+(t)$.

	Next, we refine these solutions by incorporating the mode expansions $g_{1k} = g_k e^{-i\vec{k} \cdot \vec{r}}$ and $g_{2k} = f_k e^{-i\vec{k} \cdot \vec{r}}$.  The spectral densities of each bath is defined   through
	\begin{eqnarray}
		J_1(\omega)= \sum_k |g_k|^2 \delta(\omega-\omega_{1k}),~~~J_2(\omega) =\sum_k |f_k|^2\delta(\omega-\omega_{2k}).
	\end{eqnarray}
	Phenomenologically, we assume the Gaussian cut-off for high frequencies and model $J(\omega)$ as 
	$J_1(\omega) = \alpha \omega^s e^{-\omega^2 / \Omega^2}$ and $J_2(\omega)= \beta \omega^{s'} e^{-\omega^2 / \Omega^2}$,  
	where $\alpha$ and $\beta$ represent the intrinsic coupling constants, $\Omega$ is high frequency cut-off. The exponents $s$ and $s'$
	characterize the behavior of the spectral densities and play a crucial role in determining the type of bath under consideration. for example, $0<s,s^\prime<1$ represent sub-Ohmic bath, while $s,s^\prime=1$ and $s,s^\prime>1$ define Ohmic and super-Ohmic baths respectively. 
	Therefore, we write


	 \begin{eqnarray*}
		\label{SPD}
	 \sum_k \bigg(\frac{|g_{1k}|^2}{\omega_{1k}^2} \cos [\omega_{1k} (t-\tau)] +\frac{|g_{2k}|^2}{\omega_{2k}^2} \cos [\omega_{2k} (t-\tau)]\bigg) &=& \sum_{k}\bigg(\frac{|g_k|^2}{\omega^2_{k}} \cos\omega_{k}(t-\tau)+\frac{|f_k|^2}{\omega^2_{k}} \cos\omega_{k}(t-\tau)\bigg) \nonumber  \\
	 &=& 
		  \int_{0}^{\infty}  d\omega \frac{ \big(\alpha\omega^{s}+ \beta  \omega^{s'}\big) e^{\frac{-\omega^2}{\Omega^2}}\cos\omega(t-\tau)} {\omega^2} 
	\end{eqnarray*}
		where in the second step we replaced the discrete sum over bath modes with integrals over continuous frequencies by introducing spectral densities for each bath, assuming a smooth distribution of mode frequencies and couplings. 
		 Depending on the values of  $s$ and $s'$, we can have different bath combinations. For example, when the values of $s$ and $s^\prime$ are same, we have same type i.e. either  Ohmic, sub-Ohmic or Super-Ohmic   bath at each site. The other combinations involve different values of $s$ and $s^\prime$.		 
		 
\begin{figure}[t]
   	\includegraphics[width=4cm]{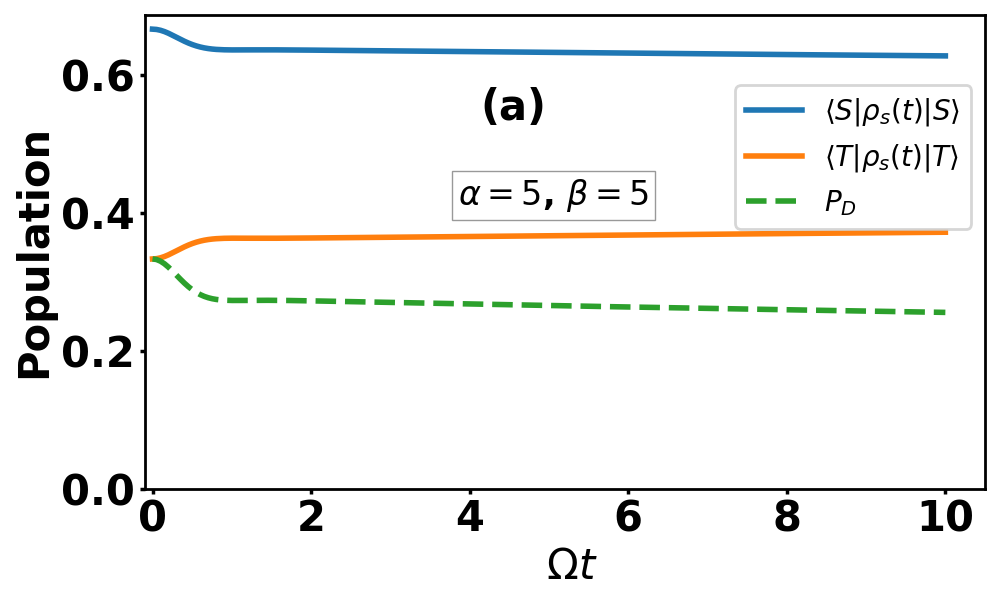} \hspace{0.1cm}
   	\includegraphics[width=4cm]{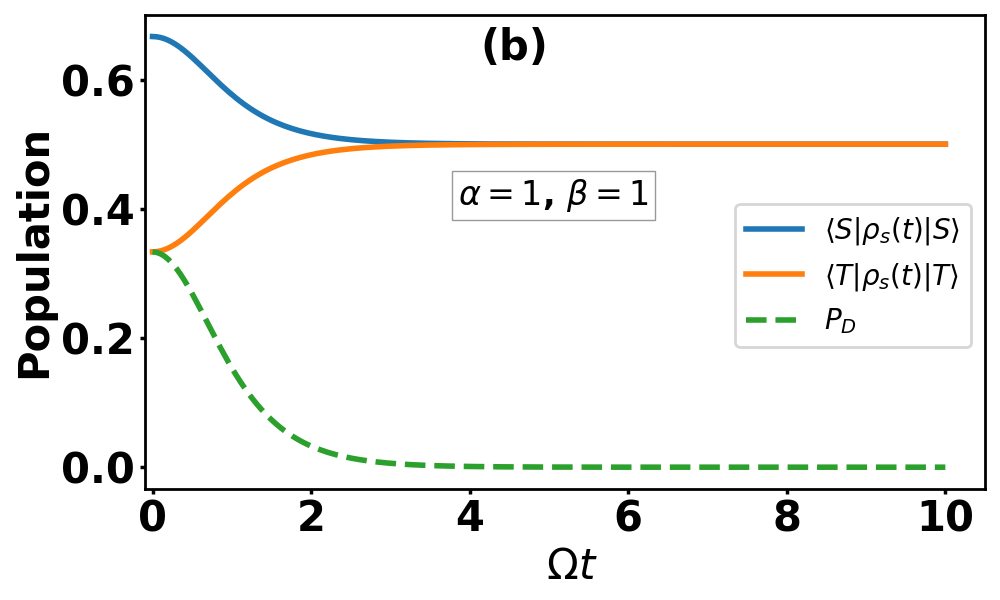} \hspace{0.1cm}
   	\includegraphics[width=4cm]{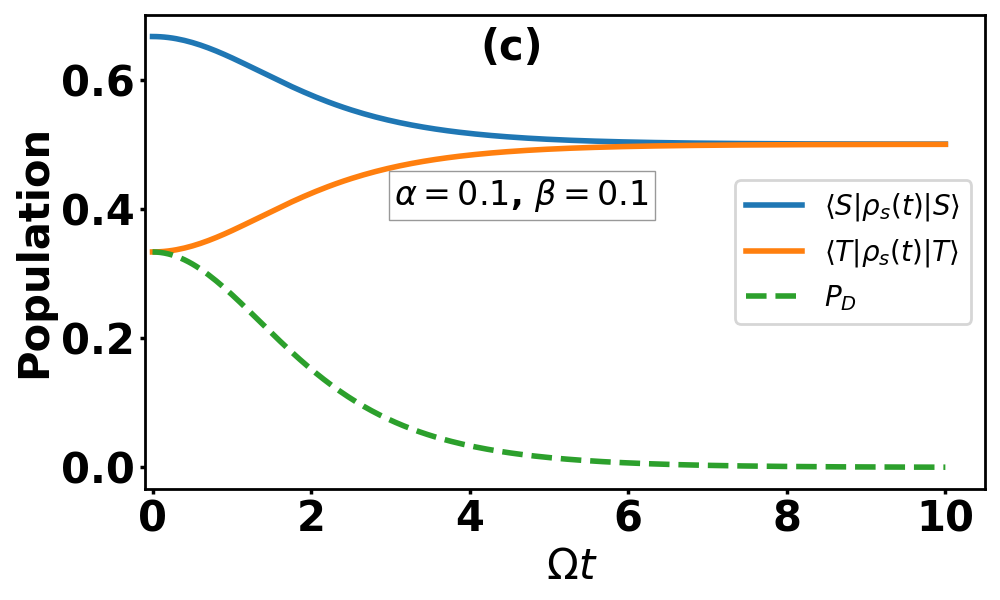} \hspace{0.1cm}
   	\includegraphics[width=4cm]{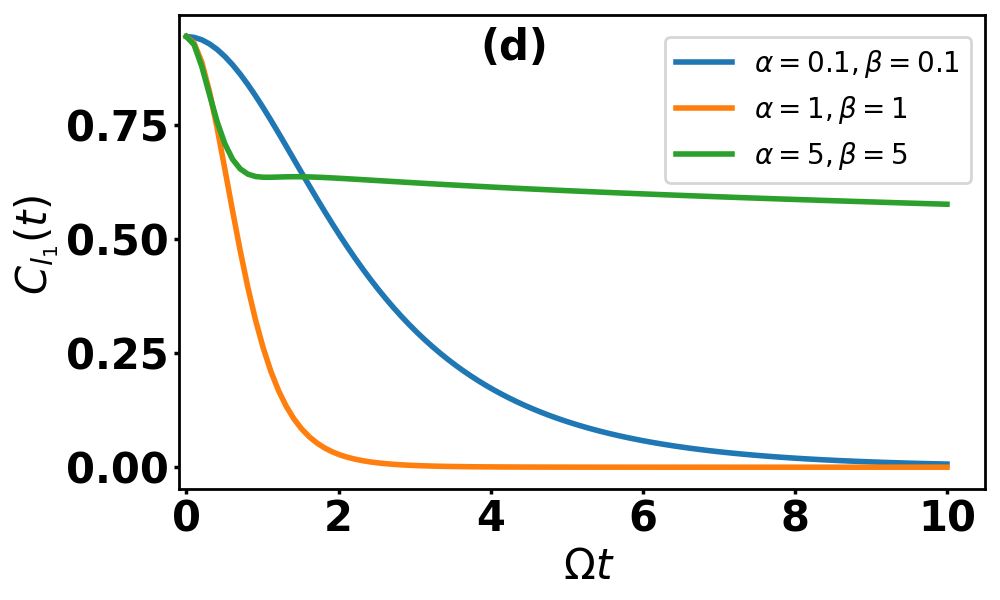}
   	\caption{
   			Time evolution of the population difference {$P_D(t)= \langle S|\rho(t)|S\rangle - \langle T|\rho(t)|T\rangle$}, populations of the $\lvert \mathcal{S} \rangle$ and $\lvert \mathcal{T} \rangle$ states, and coherence for different bath coupling strengths $\alpha$ and $\beta$, with initial state $\lvert \psi_0 \rangle = \sqrt{\frac{2}{3}}\,\lvert \mathcal{S} \rangle + \sqrt{\frac{1}{3}}\,\lvert \mathcal{T} \rangle$, for the super-Ohmic case ($s = s' = 2$). We observe a delocalization-to-localization transition: at strong couplings ($\alpha=\beta=5$), populations remain largely stationary with a persistent population difference and long-lived coherence, while at weak couplings ($\alpha=\beta=0.1$), populations equilibrate and coherence decays monotonically.
   		}
   	\label{super_super}
   \end{figure}

\begin{figure}[t]
   	\centering
   	\includegraphics[width=4cm]{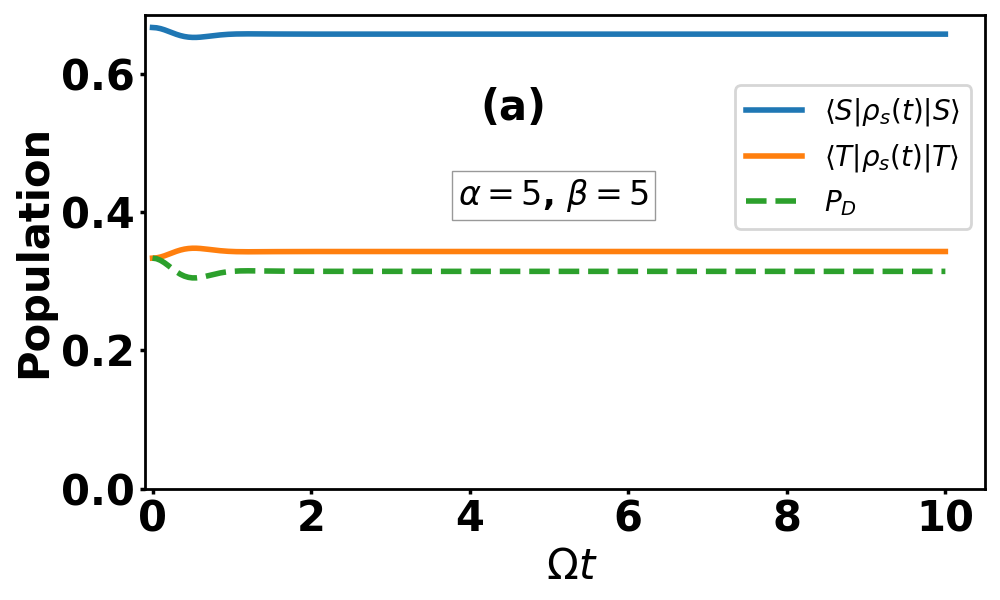} \hspace{0.1cm}
   	\includegraphics[width=4cm]{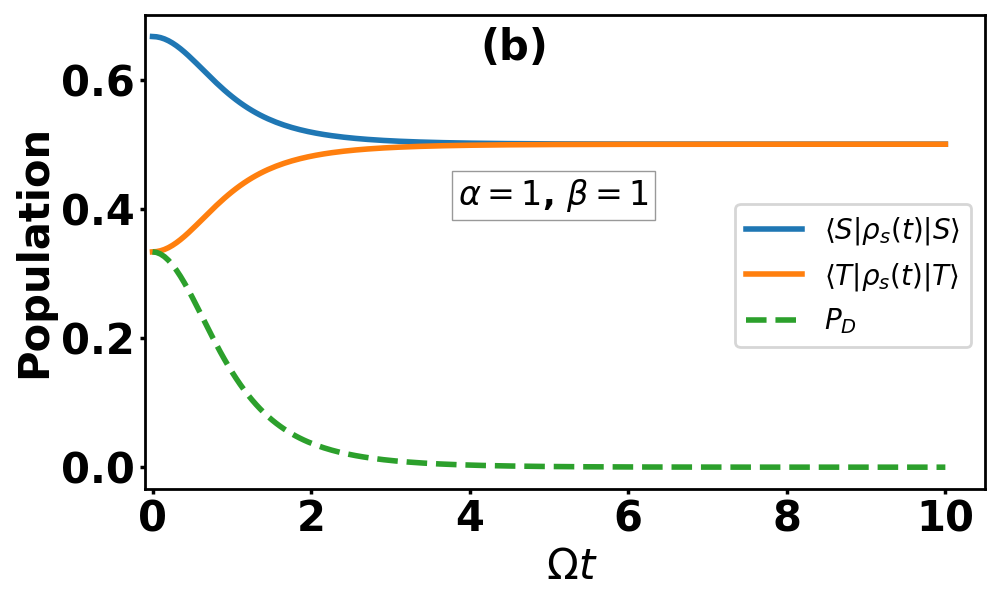} \hspace{0.1cm}
   	\includegraphics[width=4cm]{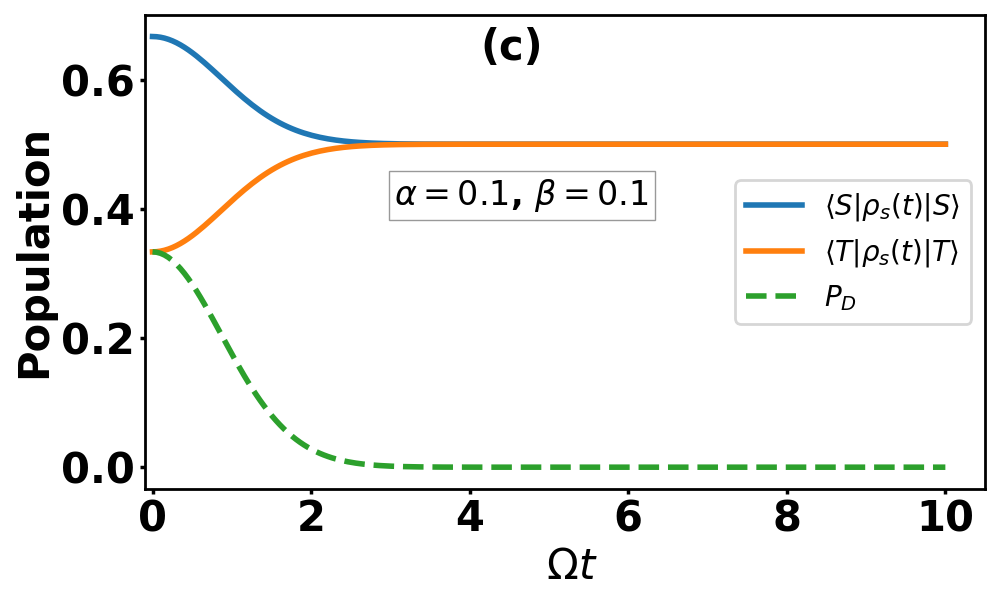} \hspace{0.1cm}
   	\includegraphics[width=4cm]{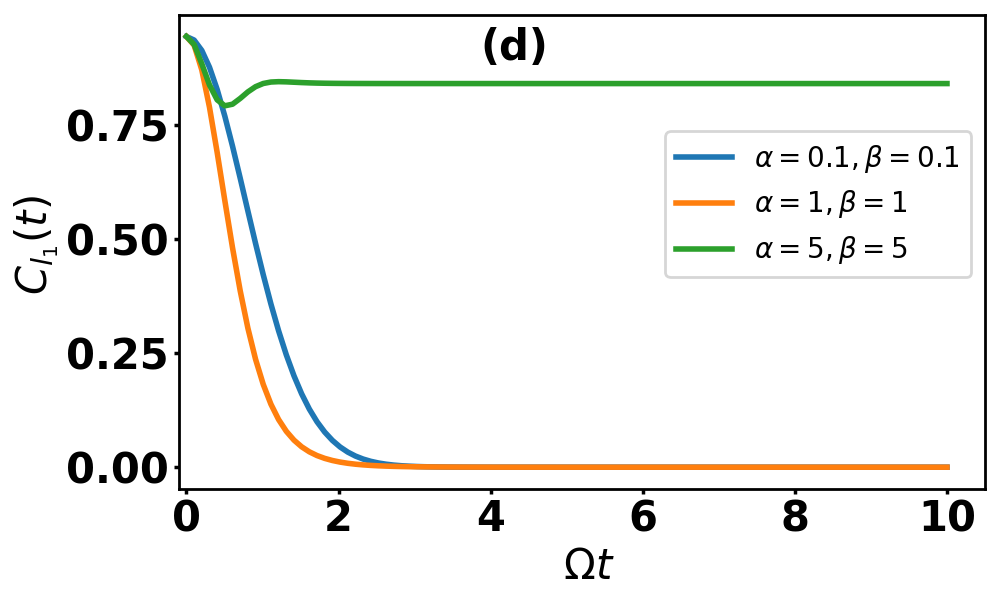}
   \caption{
  Dynamics for the combination of  Ohmic ($s = 1$) and super-Ohmic ($s' = 2$) spectral environments.
   	}
   	\label{ohmic_super}
   \end{figure}

\begin{figure}[htp]
   	\centering
   	\includegraphics[width=4cm]{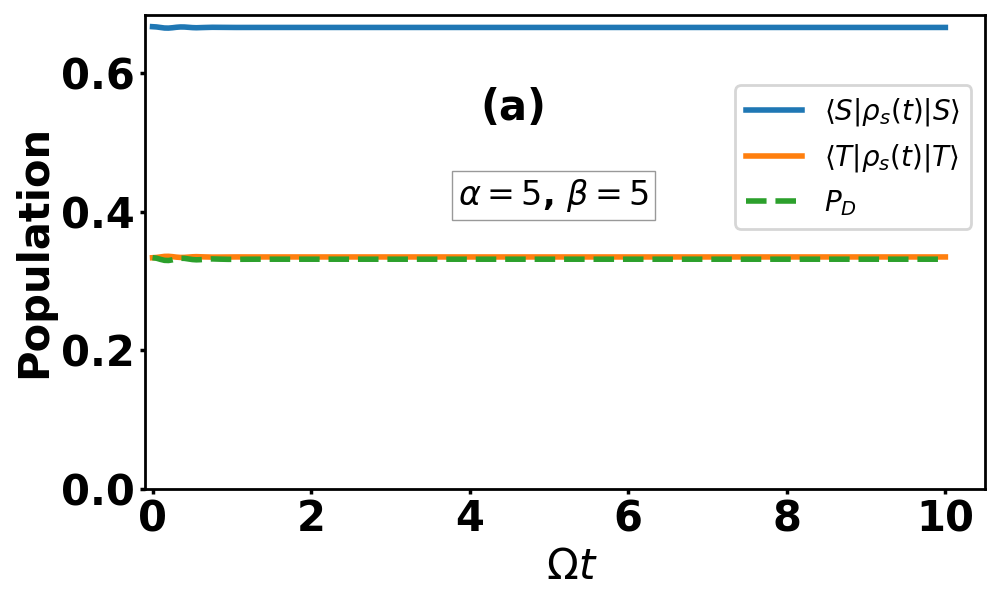} \hspace{0.1cm}
   	\includegraphics[width=4cm]{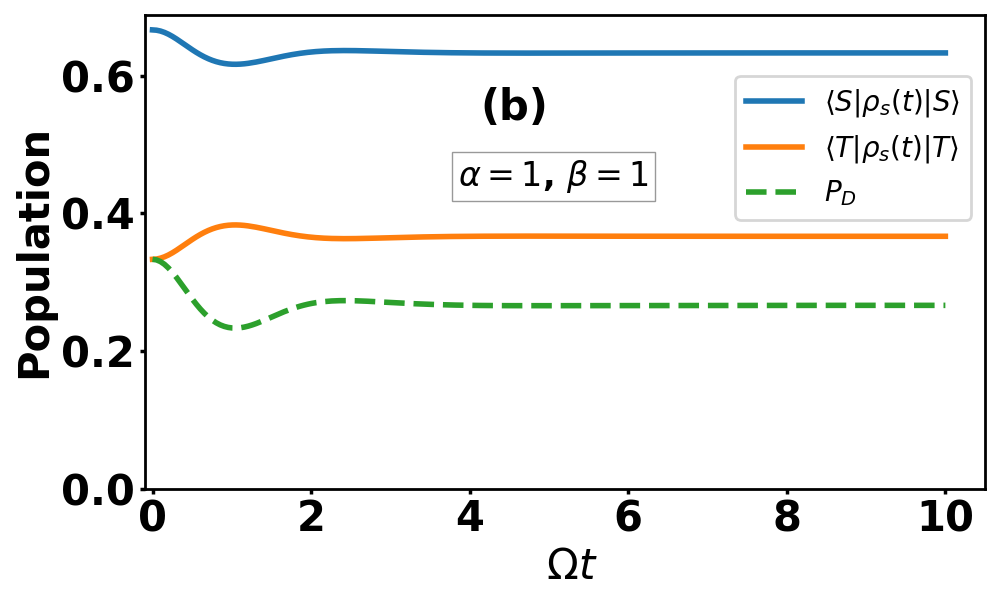} \hspace{0.1cm}
   	\includegraphics[width=4cm]{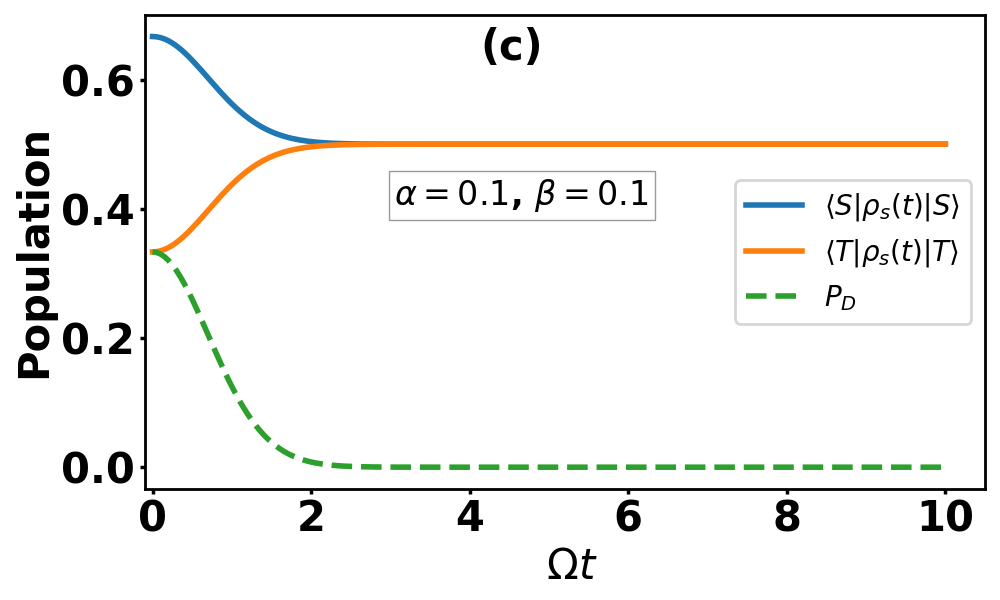} \hspace{0.1cm}
   	\includegraphics[width=4cm]{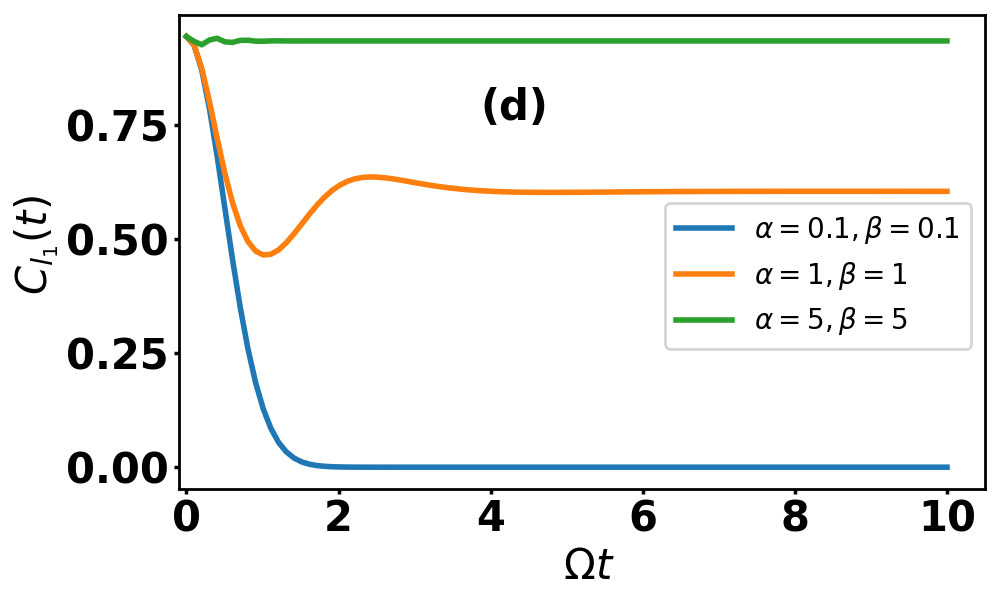}
   	\caption{
   			 Dynamics for  Sub-Ohmic ($s = 0.5$) spectral environments.
   	}
   	\label{subsub}
   \end{figure}

		 We probe  dynamics of the dressed qubit through the time evolution of  population difference  $P_D= \bra{\mathcal{S}}\rho_S\ket{\mathcal{S}}-\bra{\mathcal{T}}\rho_S\ket{\mathcal{T}}$ and coherence given by $\mathcal{C}_{l_1}= |\bra{\mathcal{S}}\rho_S\ket{\mathcal{T}}|+ |\bra{\mathcal{T}}\rho_S\ket{\mathcal{S}}|$ (see next section).  In figure \ref{super_super}, we plot $P_D(t)$ and $\mathcal{C}_{l_1}(t)$ with respect to time when both the sites are coupled to super-Ohmic bath i.e. $s=s^\prime=2$ for different values of couplings $\alpha$ and $\beta$. The initial state is chosen to be $\ket{\psi }=\sqrt{\frac{2}{3}}\,\lvert \mathcal{S} \rangle + \sqrt{\frac{1}{3}}\,\lvert \mathcal{T} \rangle $.
	     From fig. \ref{super_super}(a) to fig. \ref{super_super}(c), we observe that the system undergoes a delocalization to localization transition. In fig. \ref{super_super}(a), i.e strongly coupled case $\alpha=5,\beta=5$, the singlet and triplet populations do not decay much and remains stationary with non-zero population difference $P_D(t)$ for long times while for the weakly coupled case $\alpha=0.1,\beta=0.1$  (fig. \ref{super_super}(c)), singlet and triplet populations decay and attain a stationary value with equal populations. This is reflected in decay of coherence for these cases. For strong couplings $\alpha=5,\beta=5$, we see that system maintains coherence for long times as depicted in figure \ref{super_super}(d) while for the small values of couplings $\alpha=0.1,\beta=0.1$, coherence decays monotonically to zero. The same type of behaviour of $P_D(t)$ and $\mathcal{C}_{l_1}(t)$ is also observed when different baths couple the two sites. For example, figure \ref{ohmic_super}(a)-(d) represents the scenario when one site is coupled to Ohmic $s=1$ and other one with super-Ohmic $s^\prime=2$. However, in this case, the coherence shows non-monotonic behaviour for large couplings reflecting non-Markovianity in the dynamics. The non-Markovian effects on the dynamics are pronounced in case with $s=s^\prime=0.5$ i.e both the baths are sub-Ohmic in nature. In figure \ref{subsub}, we observe that system maintains coherence for long times with non-monotonic behaviour for strong ($\alpha=5,\beta=5$) to intermediate couplings ($\alpha=1,\beta=1$). 
		This overall trend of stronger coherence preservation and non-Markovian signatures in sub-Ohmic environments and their suppression in Ohmic and super-Ohmic combinations persists consistently across various bath configurations with different values of $s$ and $s^\prime$ (see appendix B).

\section{Non-Markovianity via coherence}
From the above section, we have observed non-monotonic behaviour of coherence, which reflects the backflow of information to the system. Measuring non-Markovianity typically involves examining the interactions between a system and its environment, monitoring the evolution of quantum states, and identifying deviations from Markovian behavior. It is important to note that the selection of a suitable measure depends on both the specific system under consideration and the characteristics of its coupling to the environment. Among the various approaches, one particularly insightful method is based on information-theoretic quantities, such as the dynamics of quantum coherence within the system. In purely Markovian dynamics, quantum coherence diminishes monotonically as the system progressively and irreversibly loses its quantum superpositions to the surrounding environment. In contrast, non-Markovian processes are distinguished by temporary increases or revivals in coherence, reflecting a backflow of information from the environment to the system. A widely adopted framework for quantifying non-Markovianity utilizes the time evolution of the $l_1$-norm of coherence under incoherent dynamical maps~\cite{Titas,chen2016quantifying,xu2020detecting}.
\begin{equation}
	C_{l_1}(\rho(t)) = \sum_{i \neq j} |\langle i | \rho(t) | j \rangle|.
\end{equation}
For Markovian evolutions, $C_{l_1}(\rho(t))$ is a monotonically non-increasing function of time. Any deviation from monotonic decay, manifested as a revival or increase in $C_{l_1}(\rho(t))$, directly signals non-Markovianity. The total amount of such coherence revival, and thus the degree of non-Markovian memory effects, can be quantified by integrating the positive increments of coherence:
\begin{equation}
	\mathcal{N} = \int_{\dot{C}_{l_1}(t) > 0} \dot{C}_{l_1}(t)\, dt,
\end{equation}
where $\dot{C}_{l_1}(t)$ is the time derivative of $C_{l_1}(\rho(t))$. A vanishing $\mathcal{N}_C$ indicates purely Markovian dynamics, while a nonzero value quantifies the cumulative memory effects present in the system.

\begin{figure}[H]
	\centering
	\includegraphics[width=5.0cm]{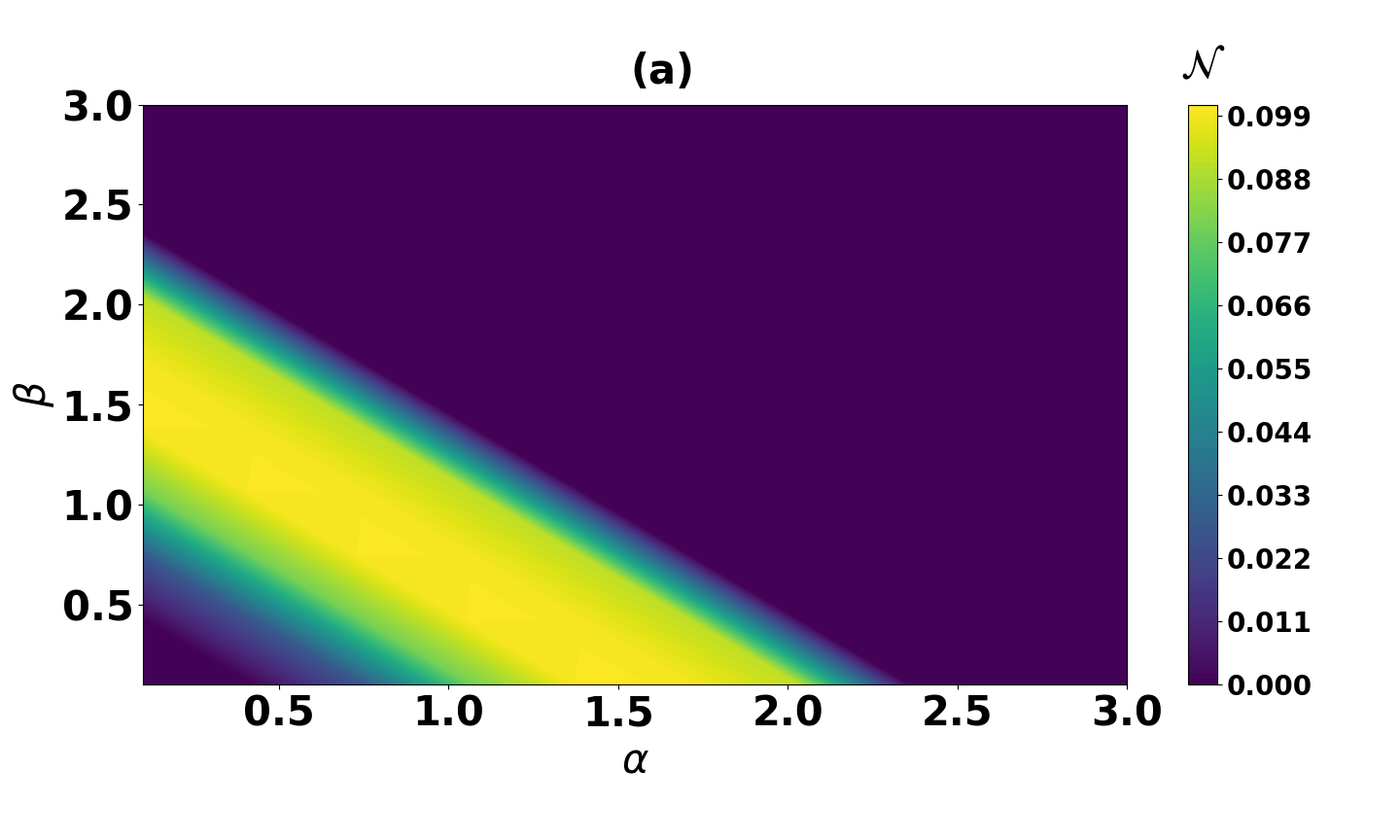} \hspace{0.1cm}
	\includegraphics[width=5.0cm]{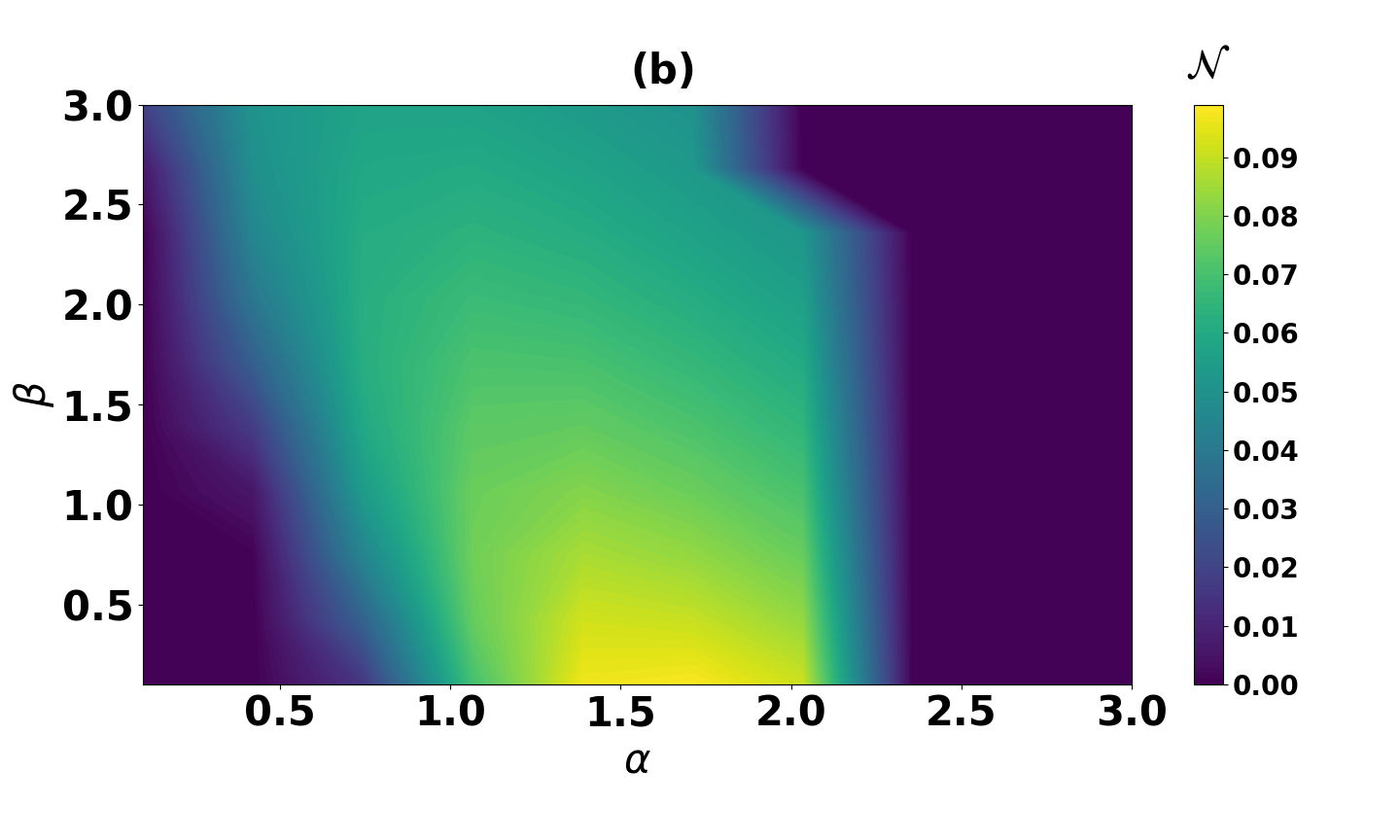}  \hspace{0.1cm}
	\includegraphics[width=5.0cm]{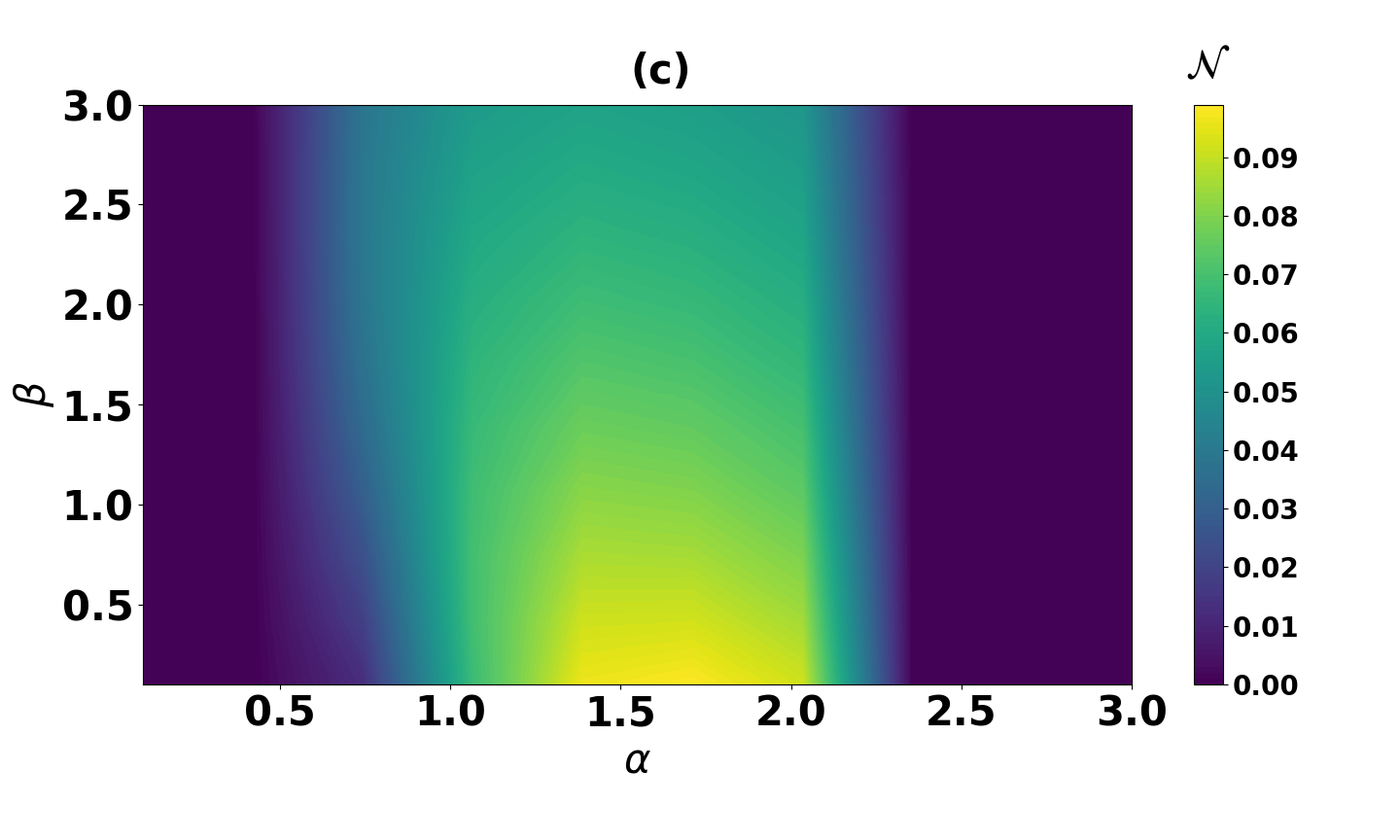} \\
	\includegraphics[width=5.0cm]{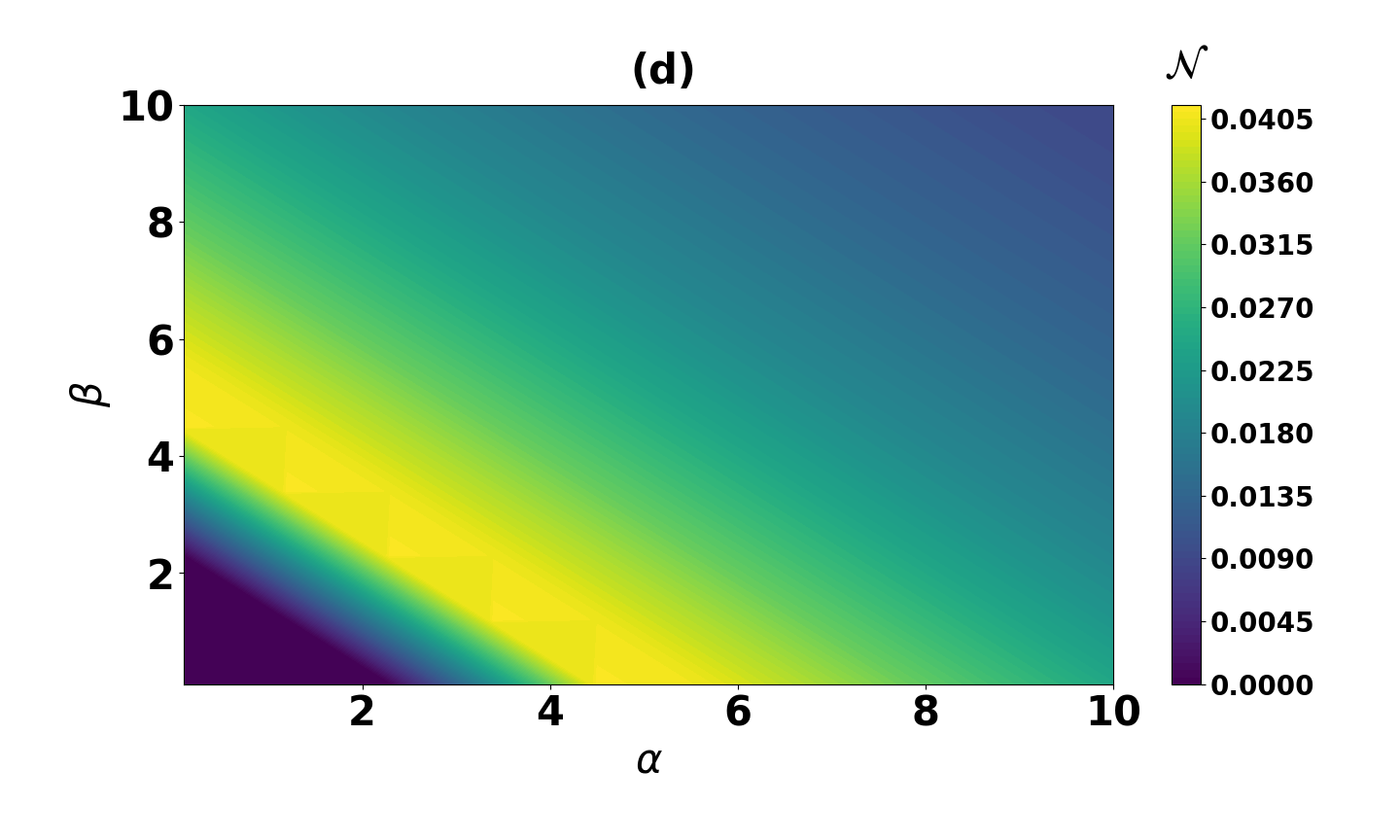}\hspace{0.1cm}
	\includegraphics[width=5.0cm]{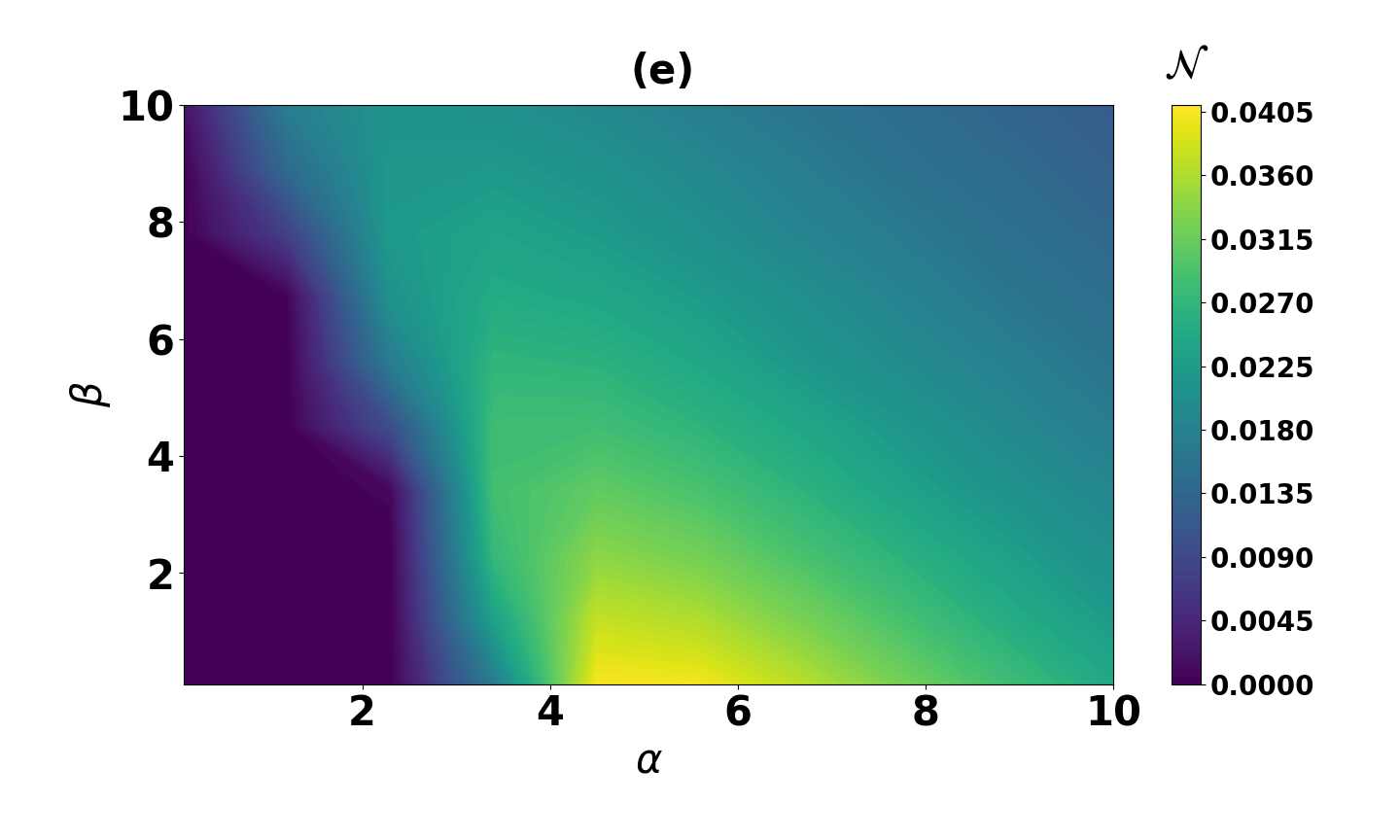}\hspace{0.1cm}
	\includegraphics[width=5.0cm]{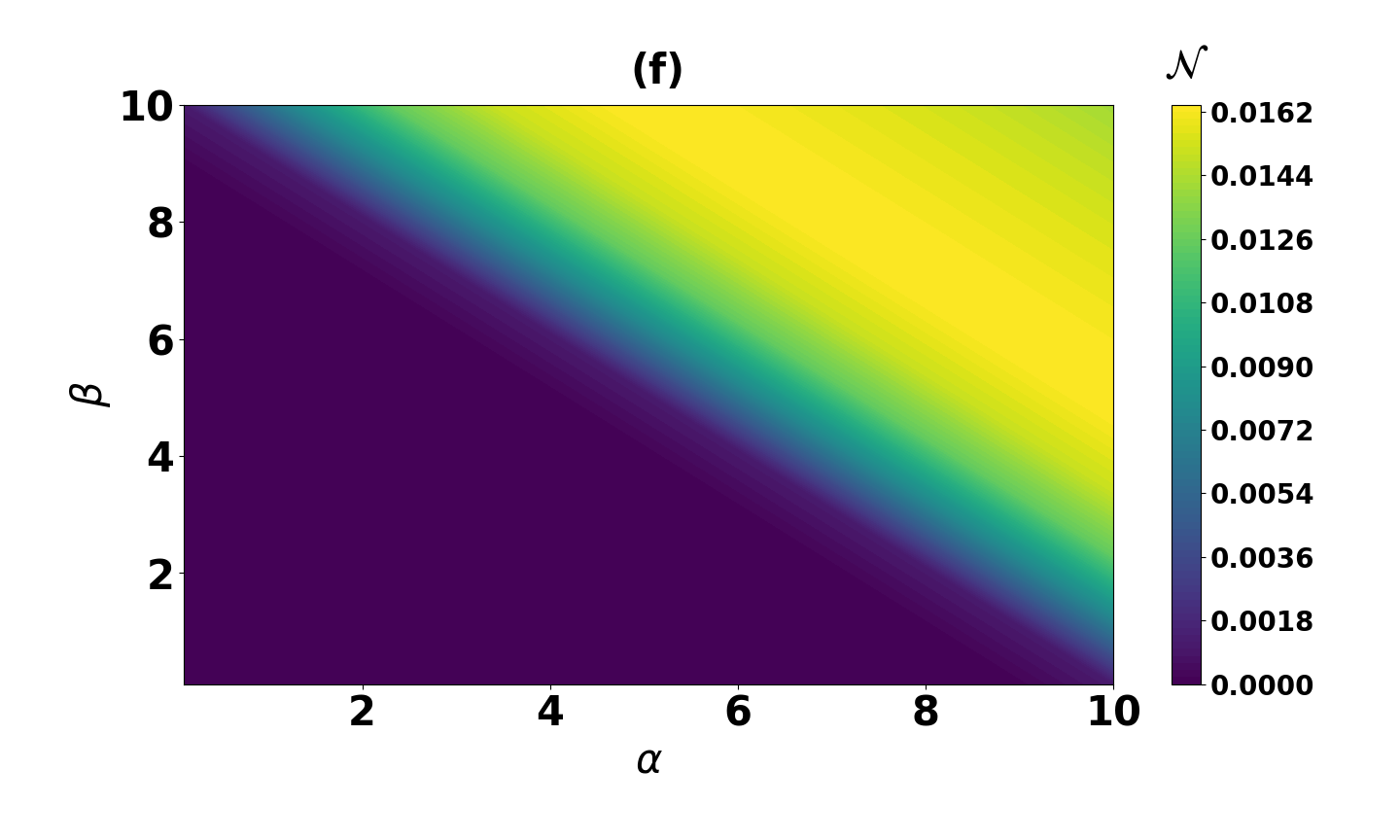}\hspace{0.1cm}
\caption{
		Non-Markovianity measure $\mathcal{N}$ as a function of coupling parameters $\alpha$ and $\beta$, for spectral exponent pairs $\{s, s'\} = \{0.5,0.5\}, \{0.5,1\}, \{0.5,2\}, \{1,1\}, \{1,2\}, \{2,2\}$. Non-Markovianity is prominent at intermediate couplings for sub-Ohmic baths, appears only at strong couplings for Ohmic–Ohmic cases, and remains negligible for mixed combinations. For super-Ohmic baths, non-Markovianity persists without coherence revivals, indicating subtle memory effects.
	}
	\label{NM}
\end{figure}

In figure \ref{NM}, we plot non-Markovianity $\mathcal{N}$ as a function of coupling parameters $\alpha$ and $\beta$. Fig. \ref{NM}(a) represent the non-Markovianity for the case of sub-Ohmic bath combinations i.e. $s=s^\prime=0.5$. Since from fig. \ref{subsub}(d), we observe that the coherence has non-monotonic behaviour for intermediate couplings $\alpha=1, \beta=1$ while for other values $\alpha=0.1,\beta=0.1$, coherence decays monotonically. This is consistent with fig \ref{NM}(a), which shows that non-zero non-Markovianity within the range of intermediate couplings. The combination of Ohmic (super-Ohmic) bath with sub-Ohmic bath has no substantial effects on non-Markovianity as seen in figure \ref{NM}(b) (fig. \ref{NM}(c)).This is consistent with the variations of coherence in figures shown in appendix B. In case of Ohmic-Ohmic combination i.e. $s=s^\prime=1$, we have non-Markovianity starting at high coupling strengths as depicted in figure \ref{NM}(c). This is also indicated in non-monotonic behaviour of coherence for strong couplings in figure \ref{ohmic_ohmic}(d). However, a peculiar case of $s=2=s^\prime$ i.e. super-Ohmic combinations, we observe no non-monotonic behaviour of coherence fig. \ref{super_super}(d) but still system posses non-Markovianity fig. \ref{NM}(f). This suggests that while coherence revivals are often indicative of information backflow, non-Markovianity can also arise through subtler mechanisms, such as temporary reductions in the overall decoherence rate or partial preservation of quantum correlations. In super-Ohmic environments, the rapid decay of bath correlations typically suppresses visible revivals, yet under certain conditions, transient memory effects can still occur without clearly observable coherence oscillations.

\section{Conclusion}
In summary, we investigated the dynamics of a qubit system modeled as a spinless fermion hoping  between two localized sites, each interacting strongly with an independent local dephasing bath. We apply the Lang-Firsov transformation  effectively making the interaction between  system and bath weaker. This makes perturbation theory applicable. In this polaron frame, utilizing the TCL master equation, we derived both the diagonal and off-diagonal components of the system's density matrix in the single-particle basis. Our results demonstrate that the system retains coherence for longer times at strong coupling values while decohering quickly for weak couplings. Notably, the time-evolved off-diagonal elements of the density matrix do not decay to zero, indicating that the qubit avoids complete localization, which is a signature of persistent quantum correlations mediated by the bath-induced renormalized interactions. This is further substantiated by the dynamics of individual populations and population differences for a wide range of spectral densities depicting combinations of baths at different sites. 

{Crucially, our investigation into non-Markovian dynamics revealed a pronounced dependence on both the bath spectral exponents $(s, s')$ and the coupling strengths $\alpha, \beta$. By systematically analyzing the coherence dynamics in the polaron frame, we identified unambiguous signatures of memory effects, including coherence revivals, non-monotonic decay, and their quantification via the non-Markovianity measure $\mathcal{N}$. Sub-Ohmic environments $(s, s' < 1)$, characterized by dense low-frequency modes and long correlation times, sustain memory effects even at modest couplings, while Ohmic $(s=1)$ and super-Ohmic $(s, s' > 1)$ baths exhibit non-Markovianity only at larger $\alpha, \beta$ due to faster correlation decay. Beyond confirming these qualitative trends, our results provide a systematic and quantitative comparison across different spectral types, establishing two broader findings: (i) the dependence of $\mathcal{N}$ on coupling strength is intrinsically non-monotonic---negligible at weak coupling, maximized at intermediate coupling where bath correlations drive information backflow and coherence revivals, and suppressed again at strong coupling due to overdamping; and (ii) non-Markovianity measures can reveal memory effects even in the absence of coherence revivals, most notably in super-Ohmic baths. This dual perspective highlights the complementary roles of coherence-based indicators and rigorous measures, showing that non-Markovianity reflects deeper structural properties of the dynamical map rather than being reducible to revival phenomena alone.}

\begin{itemize}
\item \textit{ Data availability}: No data are associated with this manuscript.
\item \textit{ Conflict of interest}: The authors declare no conflict of interest.
\end{itemize}
\begin{acknowledgments}
SB would like to thank DST, Government of India, for financial assistance through the INSPIRE fellowship (No. DST/INSPIRE Fellowship/IF210401).
\end{acknowledgments}

{\appendix
	\section{Master equation calculation}

	{In this appendix, we present a detailed derivation of the time-convolutionless master equation (\ref{ME}) for the dressed qubit. The evolution of the system's density matrix is given by}
	
	\begin{eqnarray}
		\label{NME}
		\frac{d{\rho_S}(t)}{dt}& = &- \int_0^t d \tau {\rm Tr_B}\big[   \tilde{H}_C(t)  \tilde{H}_C(\tau) {\rho}_s(t) \rho_B +{\rho}_s(t) \rho_B  \tilde{H}_C(\tau)  \tilde{H}_C(t) \nonumber \\&&~~~~~~~~~~~- \tilde{H}_C(\tau){\rho}_s (t) \rho_B \tilde{H}_C(t)- \tilde{H}_C(t){\rho}_s (t)  \rho_B \tilde{H}_C(\tau)\big]
	\end{eqnarray}
		To proceed with simplifying the master equation \ref{NME}, we begin by explicitly determining the time dependence of the interaction Hamiltonian $\tilde{H}C(t)$ in the interaction picture. This involves defining the system’s energy eigenstates as $\{|E^f_p \rangle \}$ and the energy eigenstates of the bath as$\{|\{m_{lk}\}\rangle\}$, and then we express
	\begin{eqnarray}
		\label{LFH}
		\tilde{H_C}(t) &= &e^{i H_0 t} 	\tilde{H_C} e^{-i H_0 t}\nonumber \\
		&= &e^{-i(\tilde{H_{\text{F}}}+\tilde{H_B})t }\tilde{H}_C e^{i\tilde{H_{\text{F}}}+\tilde{H_B})t }\nonumber \\
		& =&\sum_p  |E^f_{p}\rangle \langle E^f_p| \sum_{{\{m_{lk}\}}} |{\{m_{lk}\}}\rangle \langle {\{m_{lk}\}}| e^{i(\tilde{H_{\text{F}}}+\tilde{H_{\text{B}}})t }\tilde{H_C} e^{{-}i(\tilde{H_{\text{F}}}+\tilde{H_{\text{B}}})t } \sum_j  |E^f_j\rangle \langle E^f_j| \sum_{{\{n_{lk}\}}} |{\{n_{lk}\}}\rangle \langle {\{n_{lk}\}}|  \nonumber \\
		&=& \sum_{p,j} \sum_{{\{m_{lk}\}},{\{n_{lk}\}}} e^{-i[(E^f_p-E^f_j) + (\omega_{m_{lk}}-\omega_{n_{lk}})]t} |E^f_p \rangle \langle E^f_j| |{\{m_{lk}\}}\rangle \langle {\{n_{lk}\}}| \langle E^f_p| \langle {\{m_{lk}\}}|{ H}^{\prime}_I |E^f_j\rangle |{\{n_{lk}\}}\rangle.
	\end{eqnarray}
	We observe that the difference between the system energy eigenvalues is given by $\Delta E^f \equiv E^f_p-E^f_j \propto \tilde{\mathcal{J}}$ while the difference in bath energies is expressed as $\Delta E_{\text{B}} \equiv {\sum_{lk}}\omega_{m_{lk}}-\omega_{n_{lk}}={\sum_{lk}}(m_{lk}-n_{lk}) \omega_{k}$. If we now assume that $\frac{\Delta E^f}{\Delta E_{\text{B}}} <<1$\cite{Dutta,Datta,Reja,Chin,leggett1987dynamics}, which equivalently implies $\frac{\tilde{\mathcal{J}}} {\Delta E_{\text{B}}}<<1$, then the term involving $E^f_p-E^f_j$ in the exponential of equation \ref{LFH} can be safely ignored. 
	Under this condition, known as anti-adiabatic approximation, the time dependence of the system operators becomes negligible and can be safely ignored. This approximation is analogous to the Markovian approximation, where memory effects of the environment are effectively disregarded. As a result, the equation simplifies to 
	\begin{eqnarray}
		\label{ad}
		\tilde{H_C}(t)&=&  \sum_{\{m_{lk}\},\{n_{lk}\}} e^{-i (\omega_{m_{lk}}-\omega_{n_{lk}})t}  |m_{lk}\rangle \langle n_{lk}| \langle m_{lk} | \tilde{H_C} |n_{lk}\rangle = \tilde{\mathcal{J}}[ a^{\dagger}_1 a_2\mathbb{B}^{\dagger}(t)+ a^{\dagger}_2 a_1 \mathbb{B}(t)],
	\end{eqnarray}
	where $\mathbb{B}^\dagger(t)= e^{i\tilde{H}_\text{B}t}\mathbb{B}e^{-i\tilde{H}_\text{B}t} $ is the time evolved modified bath operator given by 
	
	
	\begin{eqnarray}
		\label{BB}
		\mathbb{B} (t)&=&e^{-i\tilde{H}_{\text{B}}t}\mathbb{B} e^{i\tilde{H}_{\text{B}}t}\nonumber
		\\&=&e^{-i\tilde{H}_{\text{B}}t }\big[e^{\sum_k\big(\frac{g^*_{1k}}{\omega_{1k}}a^ {\dagger}_{1k}-\frac{g^*_{2k}}{\omega_{2k}}a^ {\dagger}_{2k}\big)} e^{-\sum_k\big(\frac{g_{1k}}{\omega_{1k}}a_{1k} -\frac{g_{2k}}{\omega_{2k}}a_{2k} \big)}-1\big] e^{iH_B' t}  \nonumber
		\\&=&e^{-i\tilde{H}_{\text{B}}t}e^{\sum_k\big(\frac{g^*_{1k}}{\omega_{1k}}a^ {\dagger}_{1k}-\frac{g^*_{2k}}{\omega_{2k}}a^ {\dagger}_{2k}\big)}e^{i\tilde{H}_{\text{B}}t}e^{-i\tilde{H}_{\text{B}}t}e^{-\sum_k\big(\frac{g_{1k}}{\omega_{1k}}a_{1k} -\frac{g_{2k}}{\omega_{2k}}a_{2k} \big)}e^{i\tilde{H}_\text{B}t}-1
		\nonumber \\&=&e^{\sum_k\big(\frac{g^*_{1k}}{\omega_{1k}}a^ {\dagger}_{1k}e^{i\omega_{1k}t}-\frac{g^*_{2k}}{\omega_{2k}}a^ {\dagger}_{2k}e^{i\omega_{2k}t}\big)}e^{-\sum_k\big(\frac{g_{1k}}{\omega_{1k}}a_{1k} e^{-i\omega_{1k}t}-\frac{g_{2k}}{\omega_{2k}}a_{2k} e^{-i\omega_{2k}t}\big)}-1
	\end{eqnarray}
	{The  term $\mathrm{Tr}_B[H_I(t)H_I(\tau)\rho_S\rho_B]$ of Eq.~\ref{NME} evaluates as
	\begin{eqnarray*}
	\mathrm{Tr}_B[H_I(t)H_I(\tau)\rho_S\rho_B]&=&\tilde{\mathcal{J}}^2 \Biggl[\mathrm{Tr}_B \Big( \mathbb{B} (t) \mathbb{B}(\tau) \rho_B \Big)a_1^\dagger a_2 a_1^\dagger a_2  \rho_S+ \mathrm{Tr}_B \Big( \mathbb{B} (t) \mathbb{B} ^{\dagger}(\tau) \rho_B \Big)a_1^\dagger a_2 a_2^\dagger a_1  \rho_S\\&&+ \mathrm{Tr}_B \Big( \mathbb{B} (t) \mathbb{B} ^{\dagger}(\tau) \rho_B \Big) a_2^\dagger a_1 a_1^\dagger a_2  \rho_S+ \mathrm{Tr}_B \Big( \mathbb{B} (t) \mathbb{B} ^{\dagger}(\tau) \rho_B \Big) a_2^\dagger a_1 a_2^\dagger a_1  \rho_S\Bigg]
		\end{eqnarray*}
		and the remaining contributions follow in an analogous manner. So that, 
	the bath dynamics are encoded in the correlation functions:
	\begin{eqnarray*}
		&&C(t-\tau) = \mathrm{Tr}_B \Big( \mathbb{B} (t) \mathbb{B} ^{\dagger}(\tau) \rho_B \Big),
		\\&&C^\dagger(t - \tau) = \mathrm{Tr}_B \big( \mathbb{B} ^\dagger(t) \mathbb{B} (\tau) \rho_B \big),
		\\&&A(t - \tau) = \mathrm{Tr}_B \big( \mathbb{B} (t) \mathbb{B} (\tau) \rho_B \big),
		\\&&A^\dagger(t - \tau) = \mathrm{Tr}_B \big( \mathbb{B} ^\dagger(t) \mathbb{B} ^\dagger(\tau) \rho_B \big).
	\end{eqnarray*}
		As an illustration,we evaluate the function $A(t-\tau)$ by tracing over the vacuum bath state:
			\begin{eqnarray*}
			A(t-\tau)&=&\mathrm{Tr}_B \big( \mathbb{B} (t) \mathbb{B} (\tau) \rho_B \big)=\langle00|\mathbb{B} (t) \mathbb{B} (\tau)|00\rangle\\&=& \langle00|\Big(e^{\sum_k\big(\frac{g^*_{1k}}{\omega_{1k}}a^ {\dagger}_{1k}e^{i\omega_{1k}t}-\frac{g^*_{2k}}{\omega_{2k}}a^ {\dagger}_{2k}e^{i\omega_{2k}t}\big)}e^{-\sum_k\big(\frac{g_{1k}}{\omega_{1k}}a_{1k} e^{-i\omega_{1k}t}-\frac{g_{2k}}{\omega_{2k}}a_{2k} e^{-i\omega_{2k}t}\big)}-1\Big)\\&&\times \Big(e^{\sum_k\big(\frac{g^*_{1k}}{\omega_{1k}}a^ {\dagger}_{1k}e^{i\omega_{1k}\tau}-\frac{g^*_{2k}}{\omega_{2k}}a^ {\dagger}_{2k}e^{i\omega_{2k}\tau}\big)}e^{-\sum_k\big(\frac{g_{1k}}{\omega_{1k}}a_{1k} e^{-i\omega_{1k}\tau}-\frac{g_{2k}}{\omega_{2k}}a_{2k} e^{-i\omega_{2k}\tau}\big)}-1\Big) |00\rangle
			\\&=&\langle 00| \Big(e^{-\sum_k\big(\frac{g_{1k}}{\omega_{1k}}a_{1k} e^{-i\omega_{1k}t}-\frac{g_{2k}}{\omega_{2k}}a_{2k} e^{-i\omega_{2k}t}\big)}-1\Big)\Big(e^{\sum_k\big(\frac{g^*_{1k}}{\omega_{1k}}a^ {\dagger}_{1k}e^{i\omega_{1k}\tau}-\frac{g^*_{2k}}{\omega_{2k}}a^ {\dagger}_{2k}e^{i\omega_{2k}\tau}\big)}-1\Big)|00\rangle\\&=& \langle 00| \Big(e^{-\sum_k\big(\frac{g_{1k}}{\omega_{1k}}a_{1k} e^{-i\omega_{1k}t}-\frac{g_{2k}}{\omega_{2k}}a_{2k} e^{-i\omega_{2k}t}\big)}\Big)\Big(e^{\sum_k\big(\frac{g^*_{1k}}{\omega_{1k}}a^ {\dagger}_{1k}e^{i\omega_{1k}\tau}-\frac{g^*_{2k}}{\omega_{2k}}a^ {\dagger}_{2k}e^{i\omega_{2k}\tau}\big)}\Big)|00\rangle-1\\&=& \langle 00| \Big(e^{\sum_k\big(\frac{g^*_{1k}}{\omega_{1k}}a^ {\dagger}_{1k}e^{i\omega_{1k}\tau}-\frac{g^*_{2k}}{\omega_{2k}}a^ {\dagger}_{2k}e^{i\omega_{2k}\tau}\big)}\Big)\Big(e^{-\sum_k\big(\frac{g^*_{1k}}{\omega_{1k}}a^ {\dagger}_{1k}e^{i\omega_{1k}\tau}-\frac{g^*_{2k}}{\omega_{2k}}a^ {\dagger}_{2k}e^{i\omega_{2k}\tau}\big)}\Big)\\&&\times \Big(e^{-\sum_k\big(\frac{g_{1k}}{\omega_{1k}}a_{1k} e^{-i\omega_{1k}t}-\frac{g_{2k}}{\omega_{2k}}a_{2k} e^{-i\omega_{2k}t}\big)}\Big)\Big(e^{\sum_k\big(\frac{g^*_{1k}}{\omega_{1k}}a^ {\dagger}_{1k}e^{i\omega_{1k}\tau}-\frac{g^*_{2k}}{\omega_{2k}}a^ {\dagger}_{2k}e^{i\omega_{2k}\tau}\big)}\Big)|00\rangle-1
			\\&=& e^{-\sum_k\bigg(\frac{|g_{1k}|^2}{\omega_{1k}^2} e^{-i \omega_{1k}(t-\tau)}+\frac{|g_{2k}|^2}{\omega_{2k}^2} e^{-i \omega_{2k}(t-\tau)}\big)}-1
			\end{eqnarray*}
			}

    The other correlation functions $C(t-\tau)$, $C^\dagger(t-\tau)$, and $A^\dagger(t-\tau)$ follow from similar steps.
    Using these correlation functions in equation \ref{NME}, we obtain the master equation that describes the evolution of the system's density operator:
	\begin{eqnarray}
		\label{FE}
		\frac{d{\rho_S}(t)}{dt} &=& -\tilde{\mathcal{J} }^2 \int_0^t d \tau \bigg[(e^{\sum_k\big(\frac{|g_{1k}|^2}{\omega_{1k}^2}e^{-i\omega_{1k}(t-\tau)}+\frac{|g_{2k}|^2}{\omega_{2k}^2}e^{-i\omega_{2k}(t-\tau)}\big)}-1)\big\{c_1^{\dagger}c_2c_2^{\dagger}c_1\rho_S+c_2^{\dagger}c_1 c_1^{\dagger}c_2\rho_S\big\}\nonumber\\
		&&+(e^{\sum_k\big(\frac{|g_{1k}|^2}{\omega_{1k}^2}e^{i\omega_{1k}(t-\tau)}+\frac{|g_{2k}|^2}{\omega_{2k}^2}e^{i\omega_{2k}(t-\tau)}\big)}-1)\big\{\rho_S c_1^{\dagger}c_2c_2^{\dagger}c_1 +\rho_S c_2^{\dagger}c_1 c_1^{\dagger}c_2\}\nonumber\\
		&&-\Big\{(e^{-\sum_k\big(\frac{|g_{1k}|^2}{\omega_{1k}^2}e^{i\omega_{1k}(t-\tau)}+\frac{|g_{2k}|^2}{\omega_{2k}^2}e^{i\omega_{2k}(t-\tau)}\big)}-1)\big\{c_1^{\dagger}c_2\rho_Sc_1^{\dagger}c_2+c_2^{\dagger}c_1\rho_Sc_2^{\dagger}c_1\big\}\nonumber\\ 
		&&+(e^{\sum_k\big(\frac{|g_{1k}|^2}{\omega_{1k}^2}e^{i\omega_{1k}(t-\tau)}+\frac{|g_{2k}|^2}{\omega_{2k}^2}e^{i\omega_{2k}(t-\tau)}\big)}-1)\big\{c_1^{\dagger}c_2\rho_Sc_2^{\dagger}c_1+c_2^{\dagger}c_1\rho_Sc_1^{\dagger}c_2\big\}\Big\}\nonumber\\
		&&-\Big\{(e^{-\sum_k\big(\frac{|g_{1k}|^2}{\omega_{1k}^2}e^{-i\omega_{1k}(t-\tau)}+\frac{|g_{2k}|^2}{\omega_{2k}^2}e^{-i\omega_{2k}(t-\tau)}\big)}-1)\big\{c_1^{\dagger}c_2\rho_Sc_1^{\dagger}c_2+c_2^{\dagger}c_1\rho_Sc_2^{\dagger}c_1\big\}\nonumber\\ 
		&&+(e^{\sum_k\big(\frac{|g_{1k}|^2}{\omega_{1k}^2}e^{-i\omega_{1k}(t-\tau)}+\frac{|g_{2k}|^2}{\omega_{2k}^2}e^{-i\omega_{2k}(t-\tau)}\big)}-1)\big\{c_1^{\dagger}c_2\rho_Sc_2^{\dagger}c_1+c_2^{\dagger}c_1\rho_Sc_1^{\dagger}c_2\big\}\Big\}\bigg].
	\end{eqnarray}
	This equation \ref{FE} can be written as;
	\begin{eqnarray}
		\label{ME}
		\frac{d{\rho_S}(t)}{dt} &=&{i \zeta(t)\bigg[ [c_1^{\dagger}c_2c_2^{\dagger}c_1, \rho_S]-[\rho_S,c_2^{\dagger}c_1c_1^{\dagger}c_2]\bigg] } \nonumber \\&&+ \Gamma_+(t) \bigg[\big\{c_2^{\dagger}c_1\rho_S c_1^{\dagger}c_2-\frac{1 }{2}\rho_S c_1^{\dagger}c_2c_2^{\dagger}c_1-\frac{1 }{2}c_1^{\dagger}c_2c_2^{\dagger}c_1\rho_S \big\}\bigg] \nonumber\\&&+\Gamma_+(t) \bigg[\big\{c_1^{\dagger}c_2\rho_S c_2^{\dagger}c_1- \frac{1 }{2}\rho_S c_2^{\dagger}c_1c_1^{\dagger}c_2-\frac{1 }{2}c_2^{\dagger}c_1c_1^{\dagger}c_2\rho_S \big\}\bigg] \nonumber\\&&+\Gamma_-(t)\bigg[c_1^{\dagger}c_2\rho_S c_1^{\dagger}c_2+c_2^{\dagger}c_1\rho_S c_2^{\dagger}c_1\bigg]  
	\end{eqnarray}
	Where 
	\begin{align}
\Gamma_{\pm}(t)
&= 2\tilde{\mathcal{J}}^2 \int_0^t d\tau
\Bigg[
e^{\pm \sum_k \Big(
\frac{|g_{1k}|^2}{\omega_{1k}^2}
\cos [\omega_{1k} (t-\tau)]
+ \frac{|g_{2k}|^2}{\omega_{2k}^2}
\cos [\omega_{2k} (t-\tau)]
\Big)} \nonumber \\
&\quad \times
\cos \Bigg(
\sum_k \frac{|g_{1k}|^2}{\omega_{1k}^2}
\sin [\omega_{1k} (t-\tau)]
+ \frac{|g_{2k}|^2}{\omega_{2k}^2}
\sin [\omega_{2k} (t-\tau)]
\Bigg)
- 1
\Bigg],
\\[1ex]
\zeta(t)
&= \tilde{\mathcal{J}}^2 \int_0^t d\tau
\Bigg[
e^{\sum_k \Big(
\frac{|g_{1k}|^2}{\omega_{1k}^2}
\cos [\omega_{1k} (t-\tau)]
+ \frac{|g_{2k}|^2}{\omega_{2k}^2}
\cos [\omega_{2k} (t-\tau)]
\Big)} \nonumber \\
&\quad \times
\sin \Bigg(
\sum_k \frac{|g_{1k}|^2}{\omega_{1k}^2}
\sin [\omega_{1k} (t-\tau)]
+ \frac{|g_{2k}|^2}{\omega_{2k}^2}
\sin [\omega_{2k} (t-\tau)]
\Bigg)
\Bigg].
\end{align}

\section{Plots for different bath combinations and with different coupling values}
{Figures~\ref{fig:sub_super}--\ref{SS3} show the evolution of populations and coherence for different combinations of sub-Ohmic, Ohmic, and super-Ohmic baths, and for various values of the system-bath couplings $\alpha$ and $\beta$.}

  \begin{figure}[htp]
 	\centering
 	\includegraphics[width=4.25cm,height=4.25cm]{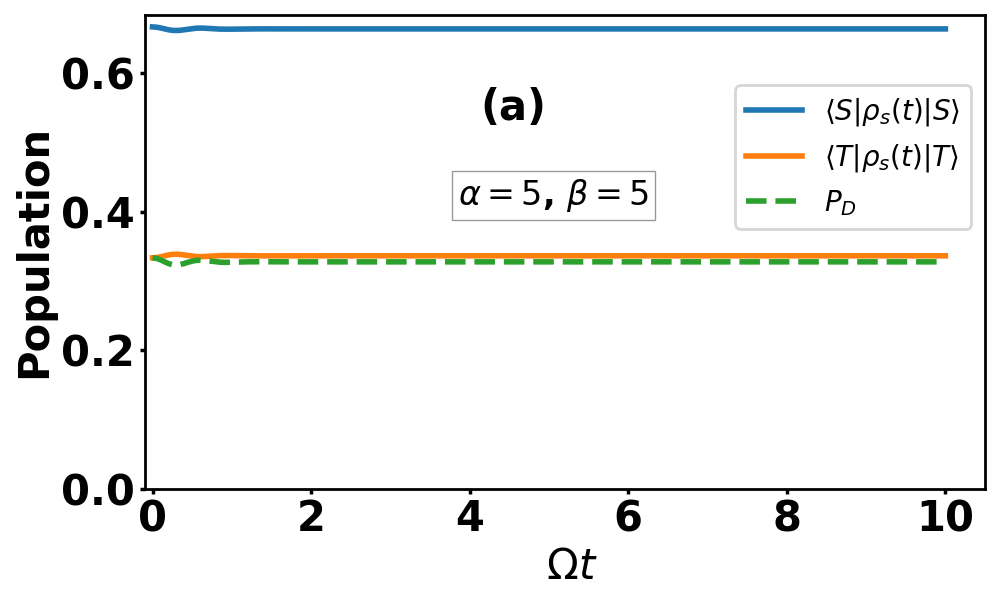} \hspace{0.1cm}
 	\includegraphics[width=4.25cm,height=4.25cm]{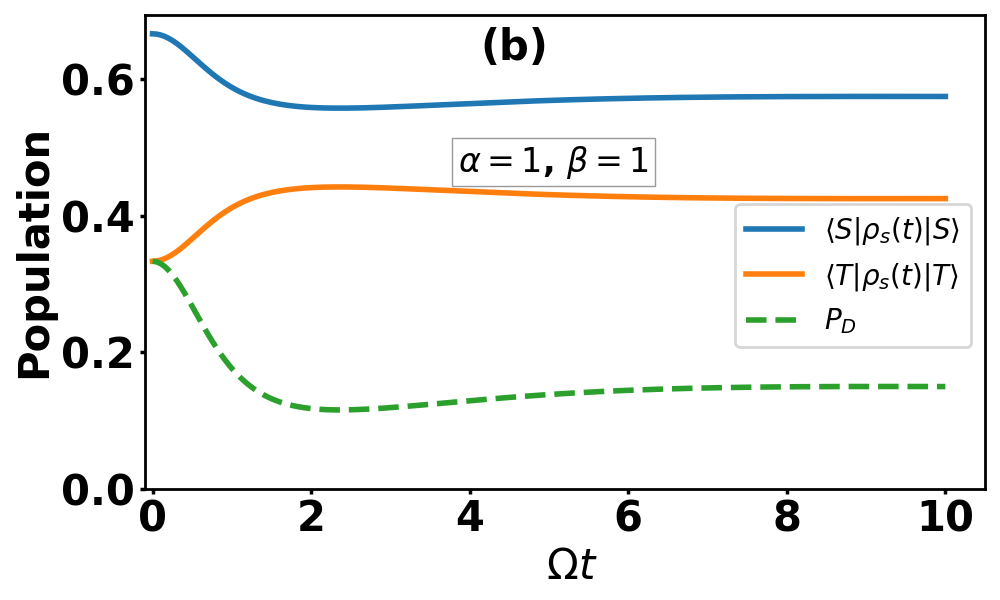} \hspace{0.1cm}
 	\includegraphics[width=4.25cm,height=4.25cm]{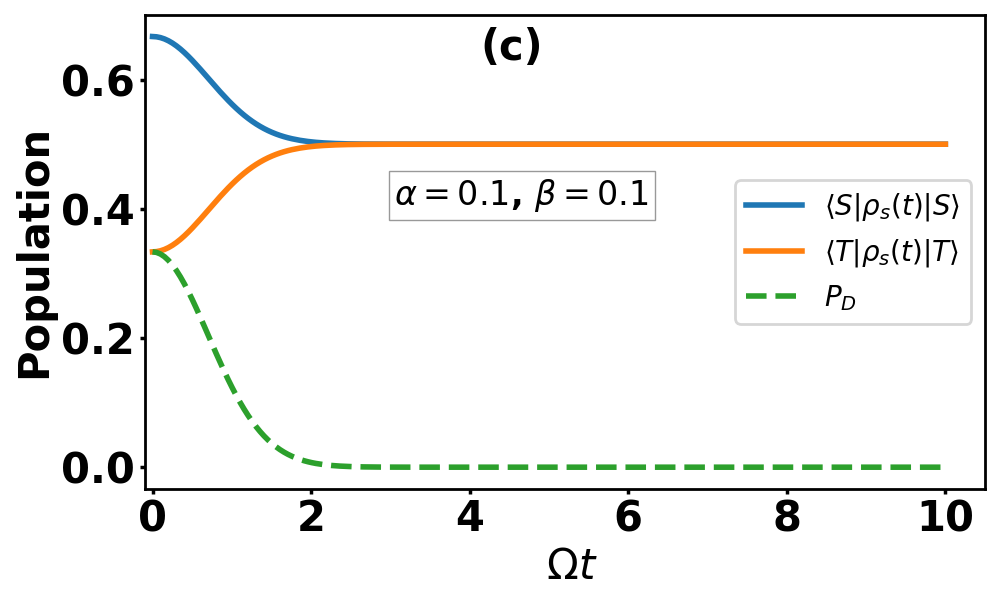} \hspace{0.1cm}
 	\includegraphics[width=4.25cm,height=4.25cm]{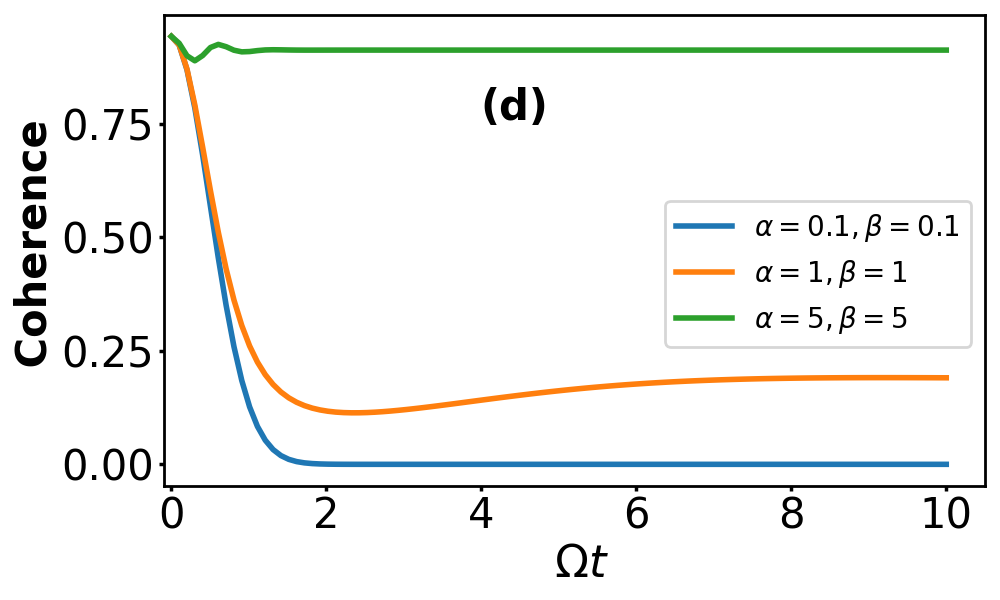}
 	\caption{  Dynamics for the combination of  sub-Ohmic ($s = 0.5$) and super-Ohmic ($s' = 2$) spectral environments.}
 	\label{fig:sub_super}
 \end{figure}
 
 \FloatBarrier
 
 \begin{figure}[htbp]
 	\centering
 	\includegraphics[width=4.25cm,height=4.25cm]{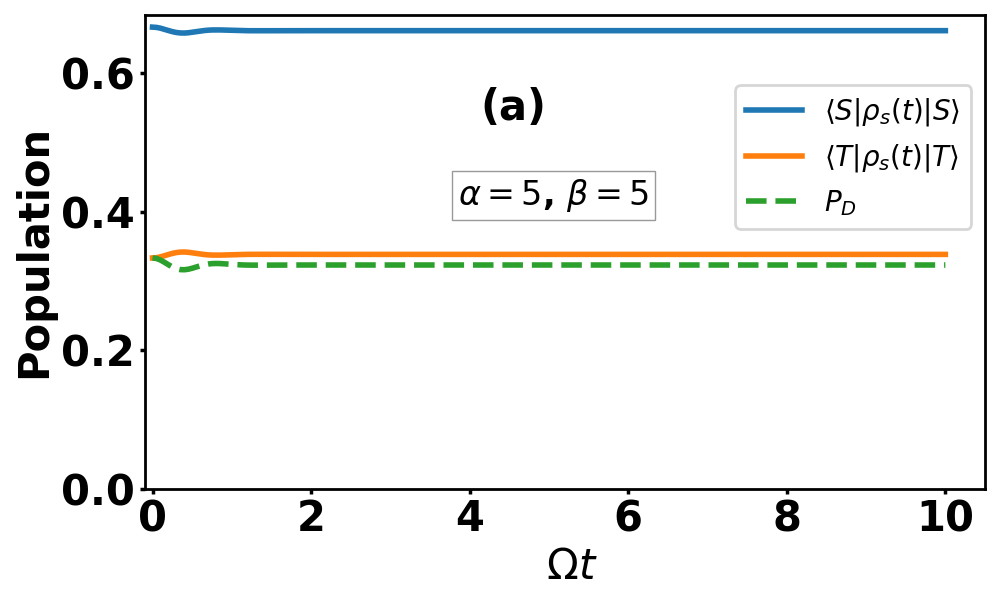} \hspace{0.1cm}
 	\includegraphics[width=4.25cm,height=4.25cm]{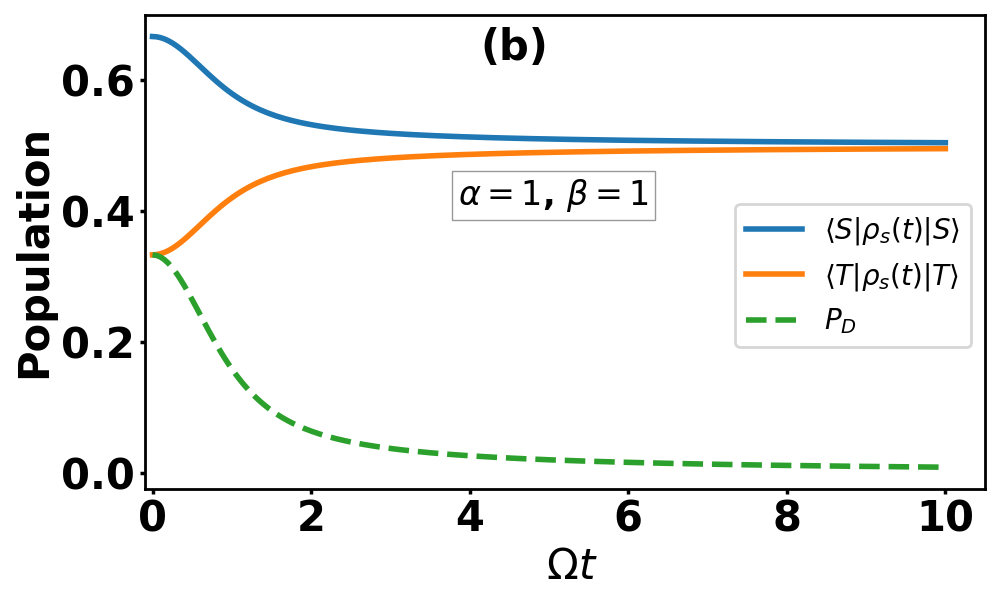} \hspace{0.1cm}
 	\includegraphics[width=4.25cm,height=4.25cm]{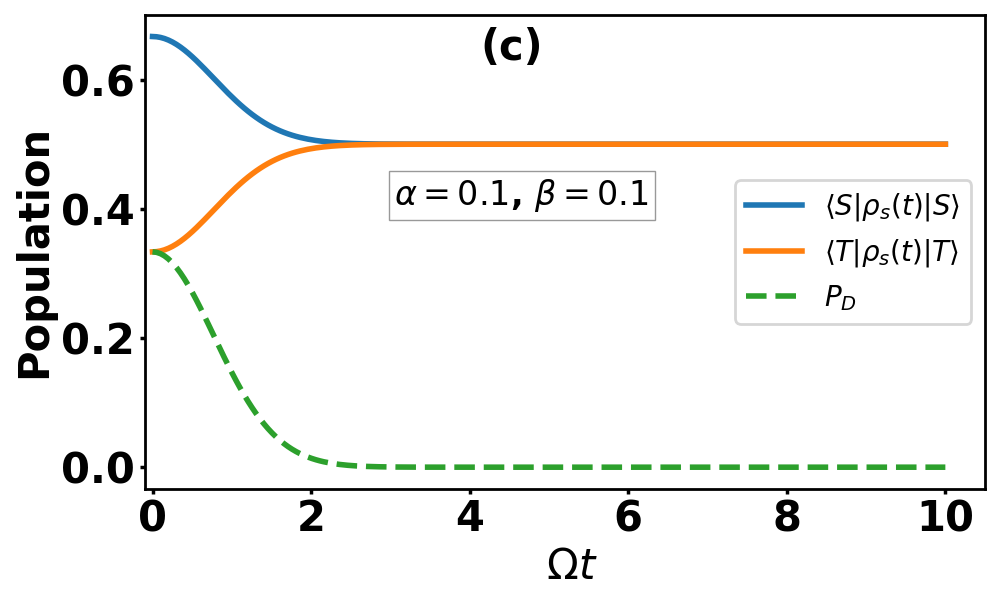} \hspace{0.1cm}
 	\includegraphics[width=4.25cm,height=4.25cm]{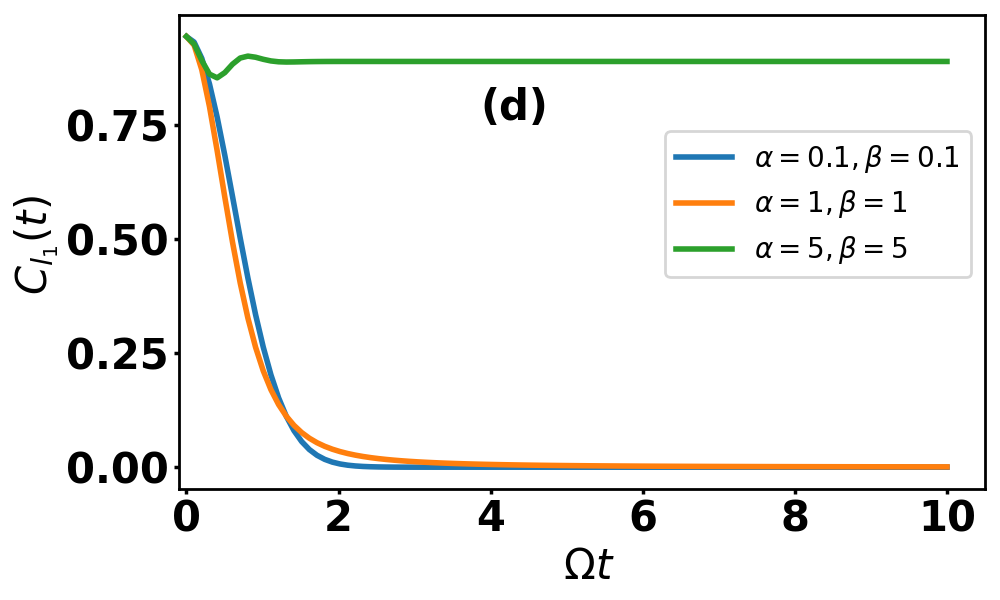}
 	\caption{  Dynamics for the Ohmic ($s = 1$) spectral environments.}
 	\label{ohmic_ohmic}
 \end{figure}
 
 \FloatBarrier
 
 
 
 \begin{figure}[htbp]
 	\centering
 	\includegraphics[width=4.25cm,height=4.25cm]{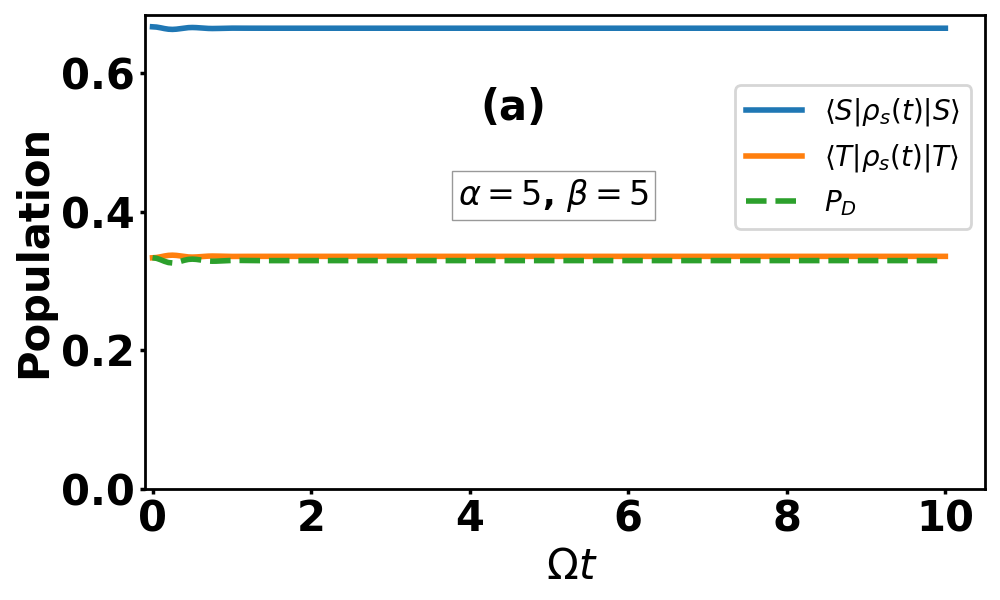} \hspace{0.1cm}
 	\includegraphics[width=4.25cm,height=4.25cm]{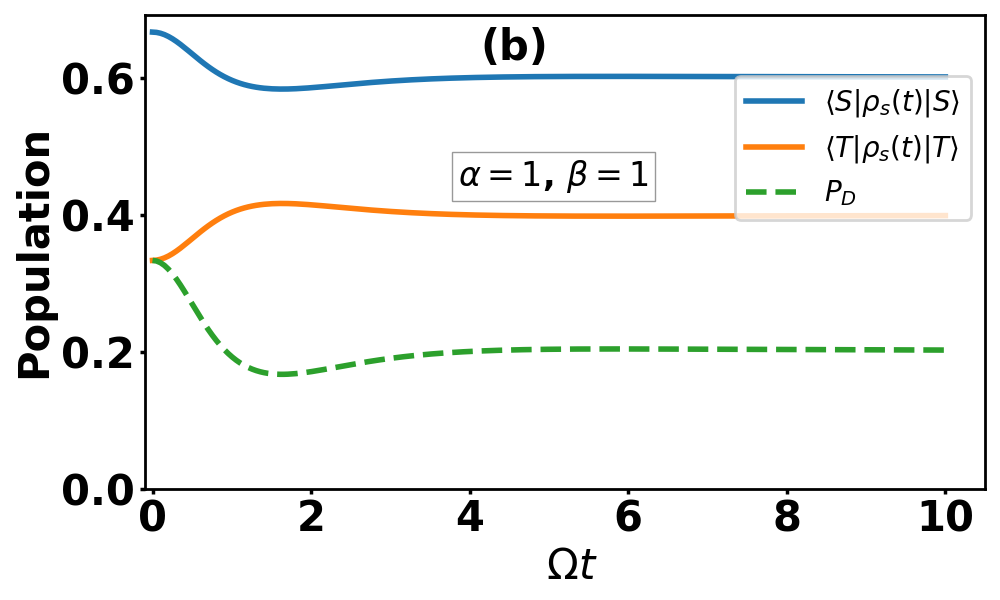} \hspace{0.1cm}
 	\includegraphics[width=4.25cm,height=4.25cm]{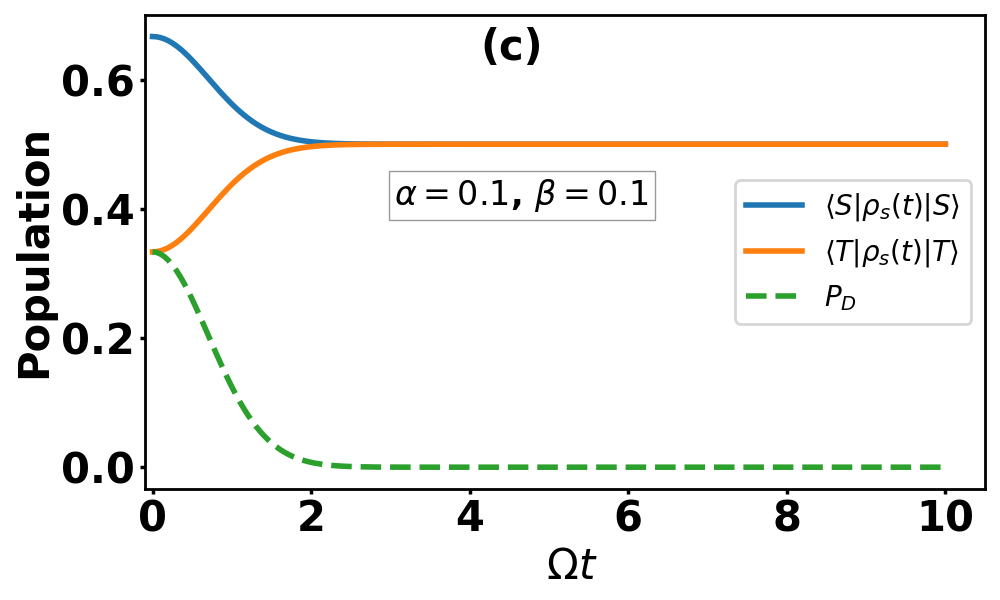} \hspace{0.1cm}
 	\includegraphics[width=4.25cm,height=4.25cm]{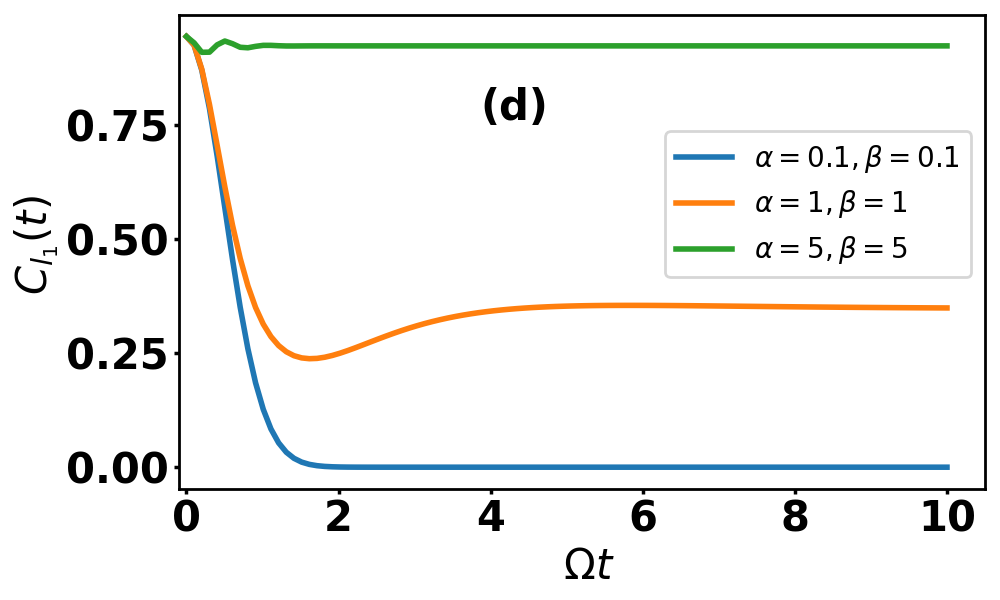}
    \caption{  Dynamics for the combination of  sub-Ohmic ($s = 0.5$) and Ohmic ($s' = 1$) spectral environments.}
 	\label{fig:sub_ohmic}
 \end{figure}
 
 \FloatBarrier
	\begin{figure}[htbp]
		\centering
		\includegraphics[width=4.25cm,height=4.25cm]{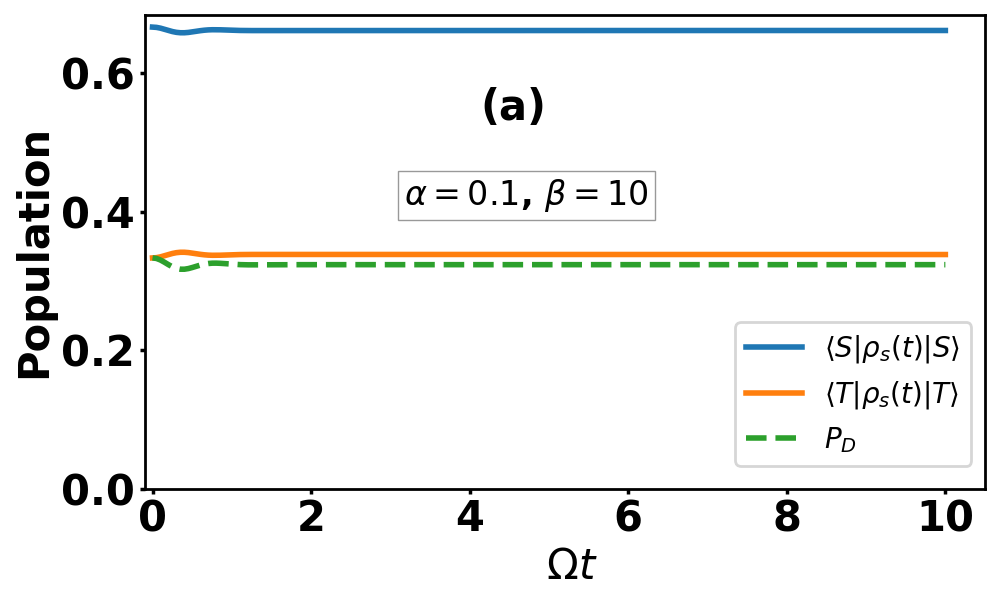} \hspace{0.1cm}
		\includegraphics[width=4.25cm,height=4.25cm]{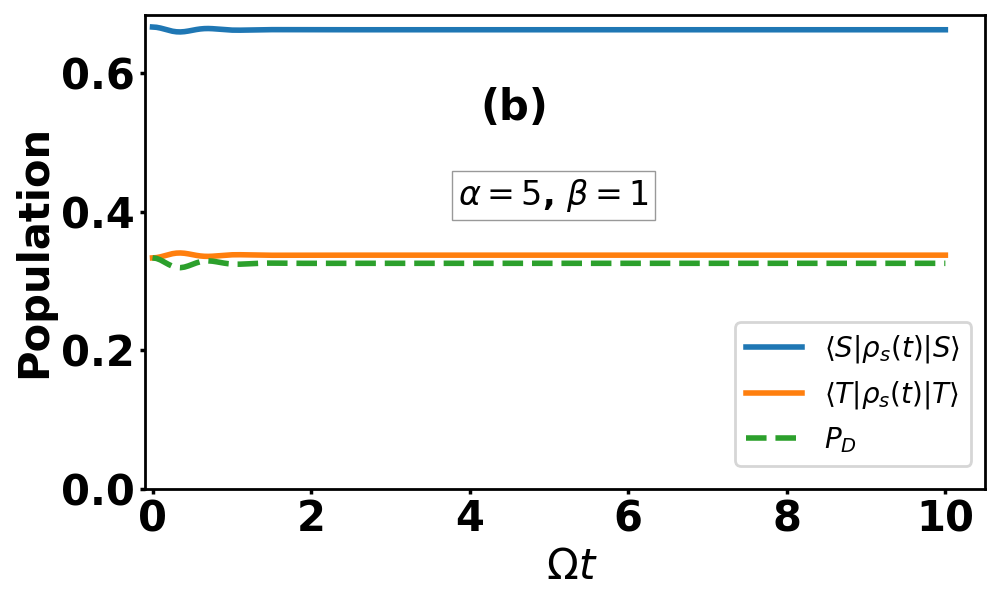} \hspace{0.1cm}
		\includegraphics[width=4.25cm,height=4.25cm]{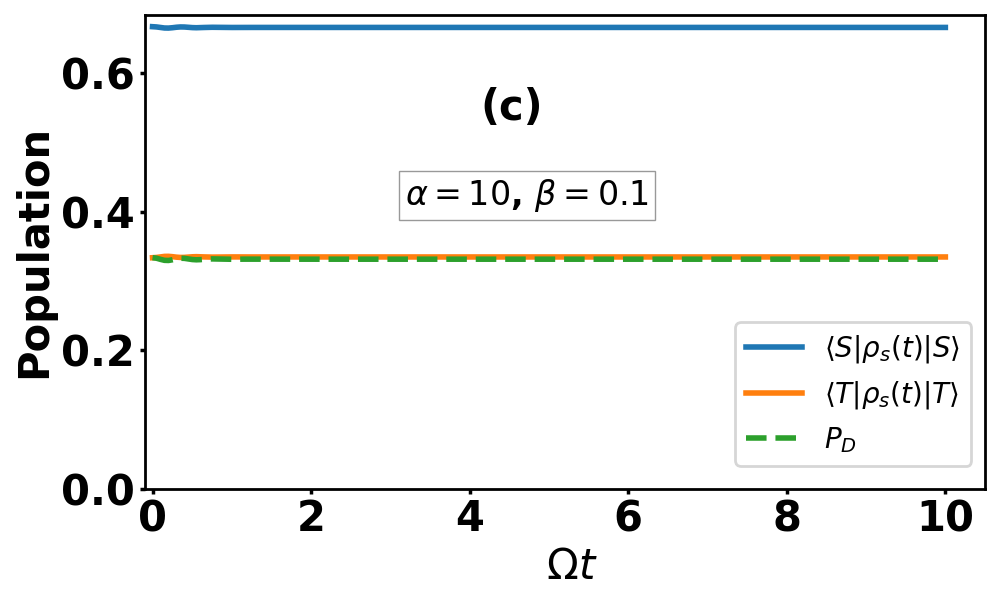} \hspace{0.1cm}
		\includegraphics[width=4.25cm,height=4.25cm]{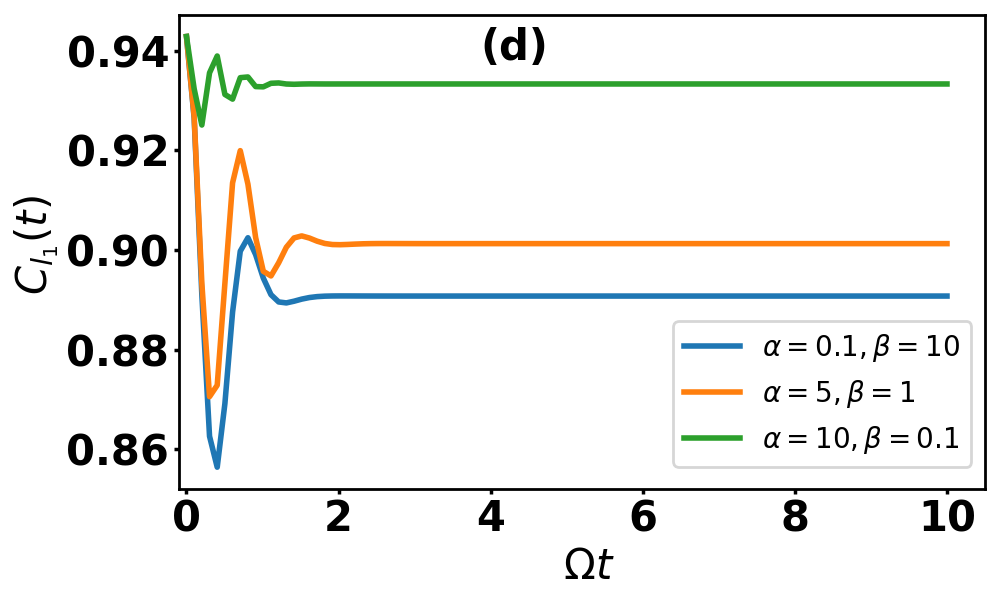} \\
		\caption{  Dynamics for the combination of  sub-Ohmic ($s = 0.5$) and Ohmic ($s' = 1$) spectral environments shown for various choices of $\alpha$ and $\beta$. 
		}
		\label{SS1}
	\end{figure}
		\begin{figure}[htbp]
		\centering
		\includegraphics[width=4.25cm,height=4.25cm]{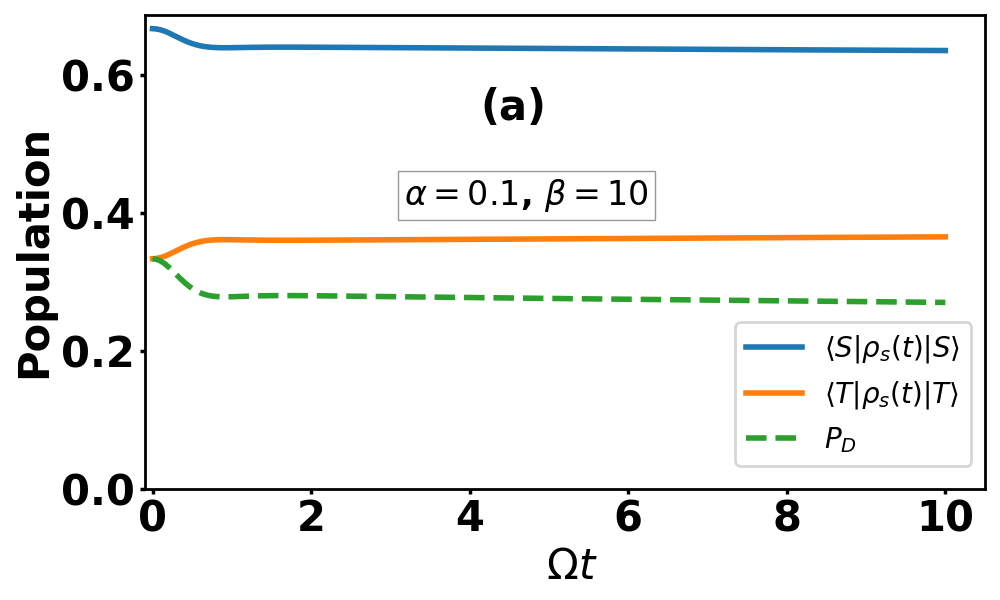} \hspace{0.1cm}
		\includegraphics[width=4.25cm,height=4.25cm]{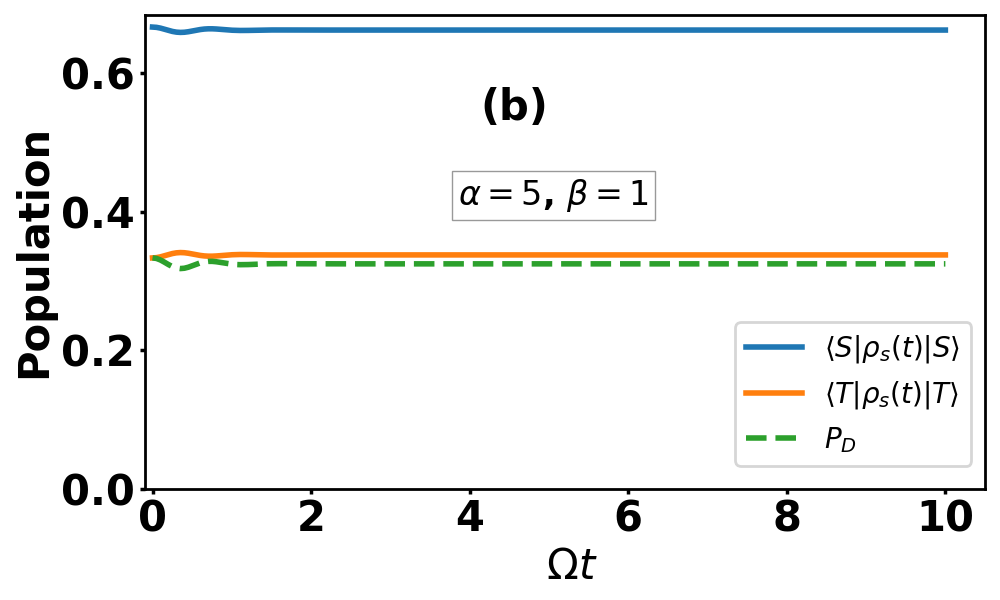} \hspace{0.1cm}
    	\includegraphics[width=4.25cm,height=4.25cm]{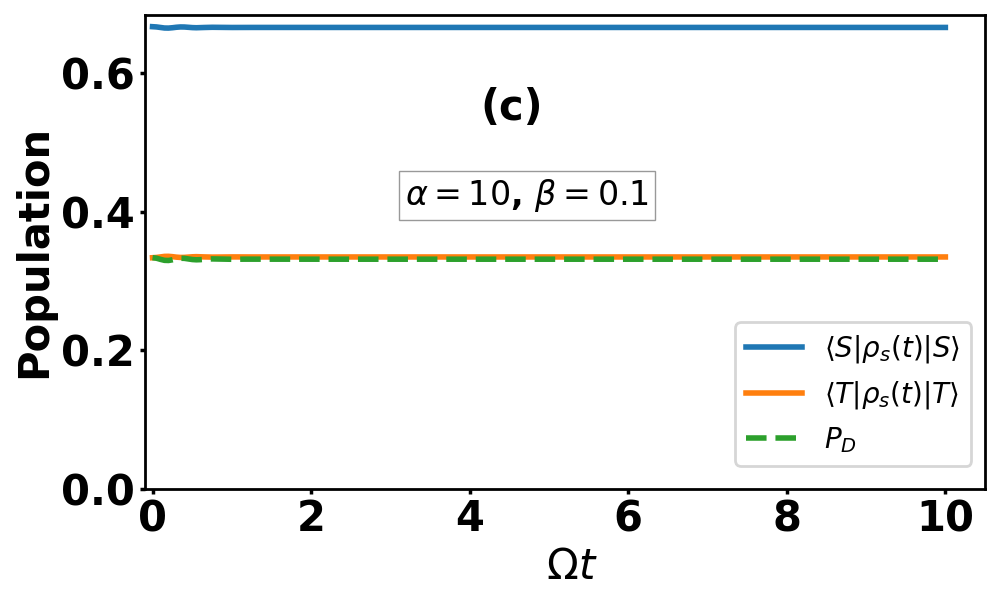} \hspace{0.1cm}
	    \includegraphics[width=4.25cm,height=4.25cm]{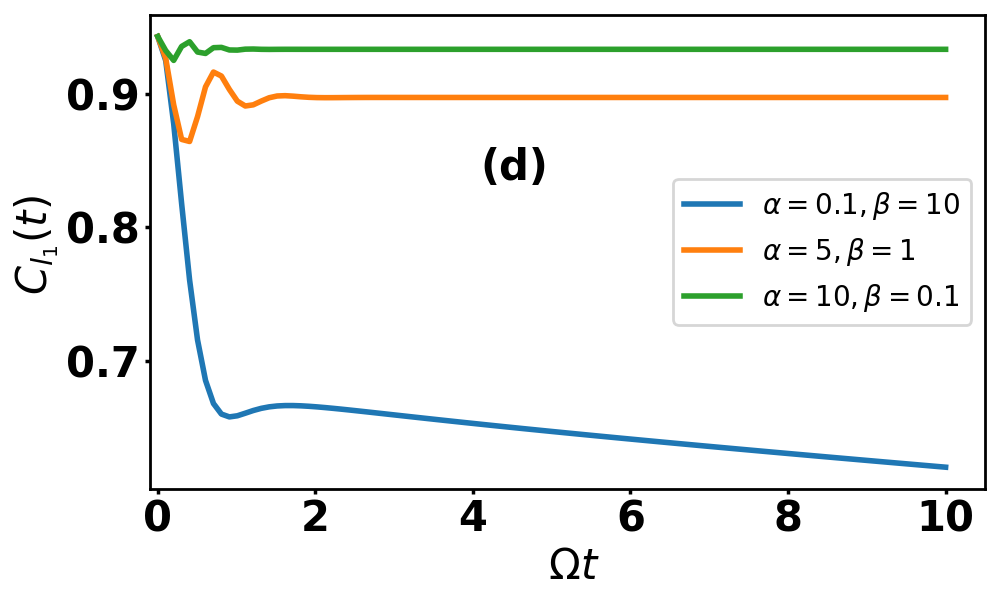}
		\caption{  Dynamics for the combination of  sub-Ohmic ($s = 0.5$) and super-Ohmic ($s' = 2$) spectral environments shown for various choices of $\alpha$ and $\beta$. 
		}
		\label{SS2}
	\end{figure}
	\FloatBarrier
		\begin{figure}[htbp]
		\centering
	\includegraphics[width=4.25cm,height=4.25cm]{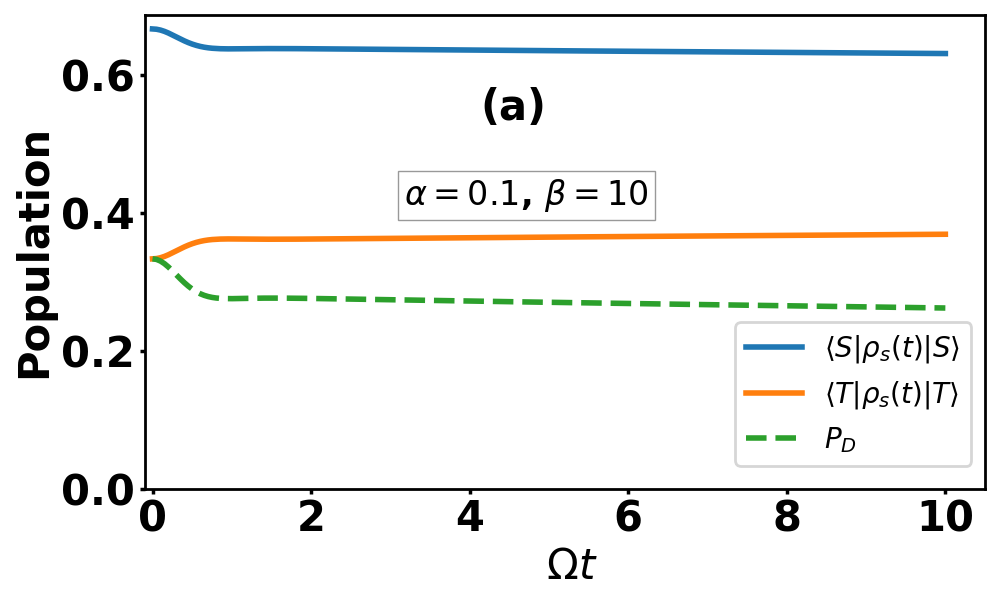} \hspace{0.1cm}
	\includegraphics[width=4.25cm,height=4.25cm]{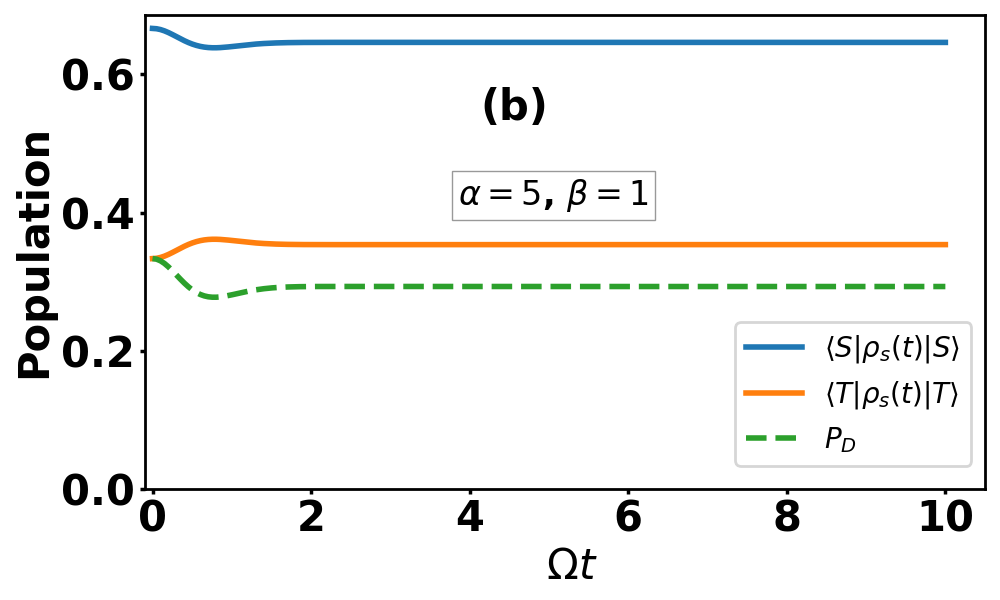}\hspace{0.1cm}
	\includegraphics[width=4.25cm,height=4.25cm]{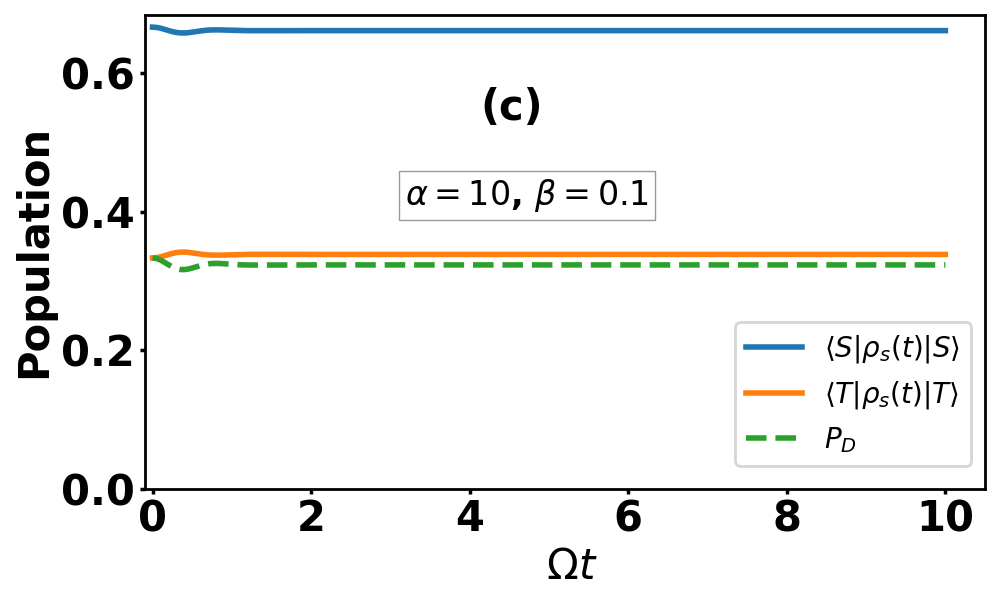}\hspace{0.1cm}
	\includegraphics[width=4.25cm,height=4.25cm]{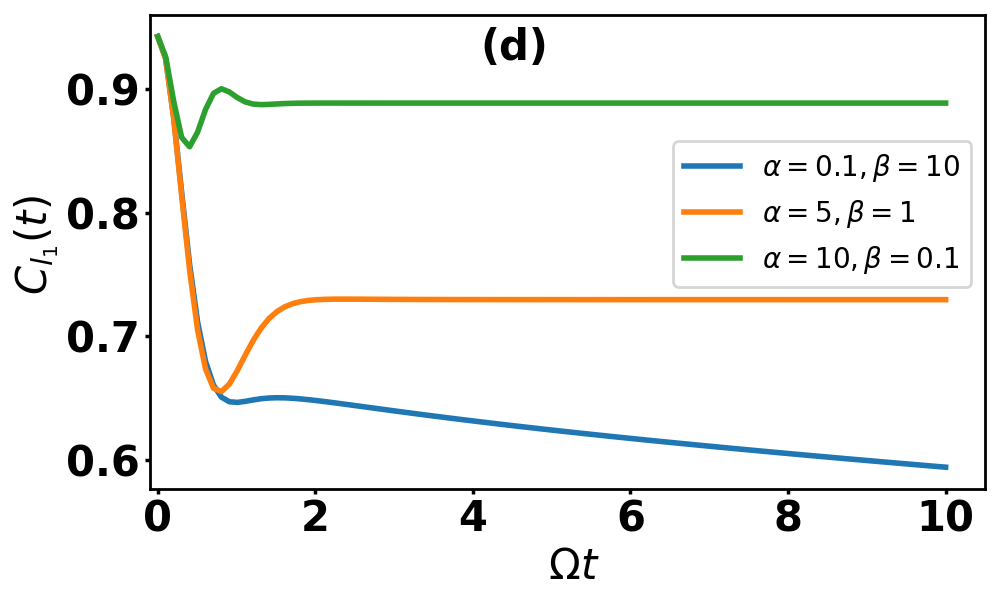}
		\caption{  Dynamics for the combination of Ohmic ($s = 1$) and super-Ohmic ($s' = 2$) spectral environments shown for various choices of $\alpha$ and $\beta$. 
		}
		\label{SS3}
	\end{figure}
	
\FloatBarrier

}
\bibliography{polaron.bib}

	
\end{document}